\documentclass{emulateapj}

\slugcomment{Accepted for publication in ApJ}

\shorttitle{Gas and dust in $z\sim1$ ULIRGs}
\shortauthors{Pope et al.}

\begin{document}

\title{Probing the Interstellar Medium of $z\sim1$ Ultra-luminous Infrared Galaxies through Interferometric Observations of CO and Spitzer Mid-infrared Spectroscopy$^{\ast}$}

\author{Alexandra Pope\altaffilmark{1},
Jeff Wagg\altaffilmark{2,3}, David Frayer\altaffilmark{4}, Lee Armus\altaffilmark{5}, Ranga-Ram Chary\altaffilmark{5}, Emanuele Daddi\altaffilmark{6}, Vandana Desai\altaffilmark{5}, Mark E.~Dickinson\altaffilmark{7}, David Elbaz\altaffilmark{6}, Jared Gabor\altaffilmark{6,8}, Allison Kirkpatrick\altaffilmark{1} 
}

\altaffiltext{1}{Department of Astronomy, University of Massachusetts, Amherst, MA 01003; pope@astro.umass.edu}
\altaffiltext{2}{European Southern Observatory, Casilla 19001, Santiago, Chile} 
\altaffiltext{3}{Cavendish Laboratory, University of Cambridge, JJ Thomson Avenue, Cambridge CB3 0HE} 
\altaffiltext{4}{National Radio Astronomy Observatory, P.O. Box 2, Green Bank, WV 24944, USA} 
\altaffiltext{5}{Spitzer Science Center, MS 220-6, California Institute of Technology, Pasadena, CA 91125} 
\altaffiltext{6}{Laboratoire AIM, CEA / DSM-CNRS-Universite ́ Paris Diderot, DAPNIA /
Service d’Astrophysique, CEA Saclay, 91191 Gif-sur-Yvette Cedex, France} 
\altaffiltext{7}{National Optical Astronomy Observatory, 950 N. Cherry Ave., Tucson, AZ, 85719}
\altaffiltext{8}{University of Arizona, 933 N. Cherry Ave., Tucson, AZ 85719}
\altaffiltext{$\ast$}{Based on observations carried out with the IRAM Plateau de Bure Interferometer. IRAM is supported by INSU/CNRS (France), MPG (Germany) and IGN (Spain).}

\begin{abstract}
We explore the relationship between gas, dust and star formation in a sample of 12 ultra-luminous infrared galaxies (ULIRGs) at high redshift compared to a similar sample of local galaxies. 
We present new CO observations and/or {\it Spitzer} mid-IR spectroscopy for 6 70$\,\mu$m-selected galaxies at $z\sim1$ in order to quantify the properties of the molecular gas reservoir, the contribution of an active galactic nuclei (AGN) to the mid-IR luminosity and the star formation efficiency (SFE=$L_{\rm{IR}}$/$L^{\prime}_{\rm{CO}}$). 
The mid-IR spectra show strong polycyclic aromatic hydrocarbon (PAH) emission and our spectral decomposition suggests that the AGN makes a minimal contribution ($<25\%$) to the mid-IR luminosity. 
The 70$\,\mu$m-selected ULIRGs which we find to be spectroscopic close pairs, are observed to have high SFE, similar to local ULIRGs and high redshift submillimeter galaxies, consistent with enhanced IR luminosity due to an ongoing major merger.
Combined with existing observations of local and high redshift ULIRGs, we further compare the PAH, IR and CO luminosities.
We show that the ratio $L_{\rm{PAH,6.2}}/L_{\rm{IR}}$ decreases with increasing IR luminosity for both local and high redshift galaxies but the trend for high redshift galaxies is shifted to higher IR luminosities; the average $L_{\rm{PAH,6.2}}/L_{\rm{IR}}$ ratio at a given $L_{\rm{IR}}$ is $\sim3$ times higher at high redshift. 
When we normalize by the molecular gas, we find this trend to be uniform for galaxies at all redshifts and that the molecular gas is correlated with the PAH dust emission. 
The similar trends seen in the [CII] to molecular gas ratios in other studies suggests that PAH emission, like [CII], continues to be a good tracer of photodissociation regions even at high redshift. Together the CO, PAH and far-IR fine structure lines should be useful for constraining the interstellar medium conditions in high redshift galaxies.

\end{abstract}


\section{Introduction}
\label{sec:intro}

The history of star formation in the Universe shows a dramatic rise from the present day to $z\sim1$ and observational evidence suggests a peak period that extends out to $z\sim3-4$ (e.g.,~Bouwens et al.~2009; Murphy et al.~2011; Magnelli et al.~2013).
This increase in the star formation rate density at high redshift can be explained by the changing gas content in galaxies; however, the mechanism for getting the gas to the galaxies is widely debated. 
Several recent theoretical studies have shown that continuous gas accretion through cold flows is sufficient to fuel the increased star formation at $z\gtrsim2$ (e.g.,~Kere{\v s} et al.~2009; Dekel et al.~2009; Dav{\'e} et al.~2010), while observations show that there is a significant contribution from an increased merger rate out to at least $z\sim1$ (e.g.,~Bridge et al.~2007; Kartaltepe et al.~2007; Conselice et al.~2009).
These two fueling modes should produce different interstellar medium (ISM) conditions, which can be probed through observations of atomic and molecular gas, and dust from small and large grains. 

One way to test these two scenarios for mass build-up is to look at the amount and distribution of gas in massive galaxies. 
Genzel et al.~(2006) observed a massive, gas-rich, rotating disk galaxy at high redshift with no evidence for interactions. Further studies of gas in UV and optically selected galaxies find large, undisturbed, rotating gas reservoirs in massive galaxies at $z\sim1$--2 (Daddi et al.~2008, 2010a; Forster Schreiber et al.~2009; Tacconi et al.~2010). 
On the other hand, $z\sim2$ submillimeter galaxies (SMGs) show disturbed, merger-like morphology in their resolved molecular CO line emission (Tacconi et al.~2008; Engel et al.~2010). 
Taking the ratio of the IR luminosity to the CO luminosity as a measure of the star formation efficiency (SFE=$L_{\rm{IR}}$/$L^{\prime}_{\rm{CO}}$\footnote{The inverse of this relation can be considered a gas depletion timescale.}, Solomon \& Vanden Bout 2005), the SMGs show higher SFE than the normal disk galaxies (Genzel et al.~2010; Daddi et al.~2010b). 
While this dichotomy is amplified by assuming different CO-H$_{2}$ conversion factors ($\alpha_{\rm{CO}}$) for starbursts and disks, it exists even when looking simply at $L_{\rm{IR}}$/$L^{\prime}_{\rm{CO}}$ (Genzel et al.~2010). 
This is expected in a merger-induced starburst where dense gas fractions are much higher (Gao \& Solomon~2004). 
The difference in average SFE between merger-induced starbursts and continuous star-forming disks has been interpreted as being due to different star formation laws; once normalized by the dynamical timescales, the dichotomy is no longer apparent (Daddi et al.~2010b; Genzel et al.~2010).

While CO line emission is tracing the molecular gas available to form stars, the mid-IR emission, specifically from polycyclic aromatic hydrocarbon (PAH) dust, is sensitive to the ongoing star formation as young stars heat the small dust grains. 
Since the advent of the {\it Spitzer Space Telescope}, mid-IR spectroscopy has been useful in quantifying the contribution from star formation and active galactic nuclei activity (AGN) to the infrared luminosity in low and high redshift galaxies since PAHs are believed to be destroyed or diluted in harsh AGN radiation fields (e.g.,~Armus et al.~2007; Sajina et al.~2007; Valiante et al.~2007; Pope et al.~2008a; Menendez-Delmestre et al.~2009; Wu et al.~2010). Surprisingly, high redshift SMGs were found to have strong PAH emission, with $L_{\rm{PAH}}/L_{\rm{IR}}$ ratios more similar to less luminous local starburst galaxies than to local ultra-luminous infrared galaxies (ULIRGs) of the same luminosity (Pope et al.~2008a). 

The lower $L_{\rm{PAH}}/L_{\rm{IR}}$ ratios for local ULIRGs compared to high redshift ULIRGS may be related to the observed deficiency in some far-IR atomic fine structure lines such as [CII] in high luminosity systems since these lines can have a significant fraction of their flux originate from photodissociation regions (PDRs, Hollenbach and Tielens 1999; Laurent et al.~2000), although there is a component which can come from diffuse, ionized gas. 
The intimate relationship between PAH and far-IR [CII] emission was demonstrated in Helou et al.~(2001) who observed nearly constant ratios of [CII] and PAH emission in a wide range of nearby star-forming galaxies. They propose that photoelectric heating of the gas (which cools via [CII] emission) is dominated by the small PAH dust grains. 
More recently, Croxall et al.~(2012) confirmed this relationship on sub-kpc scales with a resolved study of PAH and [CII] emission in two local star forming galaxies using {\it Spitzer} and {\it Herschel} data. 
In local ULIRGs, the ratio of [CII] luminosity to $L_{\rm{IR}}$ was found to be significantly lower than in less luminous galaxies; this is referred to as the [CII] deficit (e.g.~Malhotra et al.~2001).
At high redshift, far-IR fine structure lines such as [CII] have been found to be more luminous than expected, shifting the so-called [CII] deficit to higher $L_{\rm{IR}}$ (Maiolino et al.~2009; Hailey-Dunsheath et al.~2010; Stacey et al.~2010; Ivison et al.~2010; Gracia-Carpio et al.~2011, hereafter GC11). 
GC11 demonstrate that after you normalize by the molecular gas mass, galaxies at all redshifts follow the same relation, and all far-IR lines begin to show the same deficiency above a fixed star formation efficiency $L_{FIR}/M_{H_{2}}$$\sim80\,$L$_{\odot}$/M$_{\odot}$. This indicates, not surprisingly, that the molecular gas reservoir is a key quantity in understanding the conditions in the ISM and the longevity of the star formation.
GC11 further hypothesize that galaxies which show the far-IR line deficiency are more likely to be major mergers due to the apparently different ISM conditions in galaxies with the highest SFE. 
Other explanations for the [CII] deficit at high $L_{\rm{IR}}$ include low metallicity (de Breuck et al.~2011), high ionization parameter of incident radiation (Abel et al.~2009; GC11) and/or significant AGN contribution (Stacey et al.~2010; Sargsyan et al.~2012; Wagg et al.~2012). 

In this paper, we explore the relationships between the dust emission, the molecular gas reservoir and the star formation activity at high redshift compared to those in local galaxies. We present new IRAM PdBI observations of CO line emission in 70$\,\mu$m-selected ULIRGs at $z\sim1$, extending the redshift coverage of current CO observations focussed on SMGs at $z>2$ and normal disks at $z\sim1.5$. We find that the counterparts to half our 70$\,\mu$m-selected sources are spectroscopic close pairs; at these close distances it is likely that these are interacting systems. We compare the SFE of these close pair systems to what is expected for isolated disk galaxies and major mergers. 
We also present new {\it Spitzer} mid-IR spectroscopy observations which are used to quantify the star formation activity by decomposing the PAH and continuum emission. 
We have assembled a sample of galaxies from the literature which have PAH, CO and IR luminosity measurements; this includes our 70$\,\mu$m-selected galaxies, SMGs and near-IR selected $BzK$ star-forming galaxies at high redshift, and local ULIRGs. We explore how the $L_{\rm{PAH}}/L_{\rm{IR}}$ ratio changes with redshift and IR luminosity, and show that the relationship between these parameters simplifies when the molecular gas is brought into the equation. 

The paper is organized as follows. Section 2 presents our new observations with IRAM PdBI and {\it Spitzer} IRS and Section 3 summarizes the literature data used in our analysis. Section 4 summarizes our main results which we discuss further in Section 5. We recap our main conclusions in Section 6. 
Throughout this paper we assume a standard cosmology with $H_{0}=71\,\rm{km}\,\rm{s}^{-1}\,\rm{Mpc}^{-1}$, $\Omega_{\rm{M}}=0.27$ and $\Omega_{\Lambda}=0.73$ (Spergel et al.~2007).

\section{New observations}
\label{sec:samp}

High redshift SMGs are found to have much stronger PAH emission than expected compared to local ULIRGs of similar luminosities (Pope et al.~2008a). However, SMGs are only a subset of the high redshift ULIRG population, specifically they are selected based on their cold dust emission. To provide a complementary sample of high redshift ULIRGs, we aim to study the gas and dust in galaxies selected on the short wavelength side of the far-IR dust peak, specifically 70$\,\mu$m. 

Our targets for CO observations and mid-IR spectroscopy are selected at 70$\,\mu$m in the {\it Spitzer} Far-Infrared Deep Extragalactic Legacy survey (FIDEL, PI M.~Dickinson) of GOODS-N. We match the sources with a $>3\sigma$ detection at 70$\,\mu$m from the Magnelli et al.~(2011) prior based catalog to the available spectroscopic redshifts from Wirth et al.~(2004) and Barger et al.~(2008). Of the 146 70$\,\mu$m-selected galaxies within the coverage of the spectroscopic surveys, 70\% have a spectroscopic redshift $z<1$, 20\% have a spectroscopic redshift $z\ge1$ and the remaining 10\% have no available spectroscopic redshift. We preferentially select galaxies with optical spectroscopic redshifts from $z=1$--1.5 to fill the relatively unexplored redshift range from previous CO studies. These targets are listed in Table~\ref{tab:sample}; three of these targets were observed with IRAM PdBI and all six were observed with {\it Spitzer} IRS. For the IRS observations, we require $S_{24}>300\,\mu$Jy and no obvious blending in the 24$\,\mu$m image. This 24$\,\mu$m flux limit is significantly lower than many early IRS observations which were biased towards more AGN dominated sources (e.g.~Sajina et al.~2007). 
Besides these constraints, the sources for which we obtained IRS spectra and CO observations should be representative of the sample of all 70$\,\mu$m selected galaxies with $z>1$. As we discuss more later in the paper, three of these 70$\,\mu$m-selected galaxies are close pairs consisting of two optical galaxies with consistent spectroscopic redshifts. We stress that our targets were not pre-selected to be close pairs suggesting that this is a common characteristic of 70$\,\mu$m-selected galaxies (see also Kartaltepe et al.~2010). In each pair, the optical galaxies are close enough that we do not individually resolve them at 24$\,\mu$m. Therefore, our IR photometry and spectroscopy measures each whole pair system.  

We estimate the total IR luminosities (8--1000$\,\mu$m) of this sample using all the available IR spectroscopy and photometry; the {\it Spitzer} IRS spectra, {\it Spitzer} MIPS 24 and 70$\,\mu$m photometry, and {\it Herschel} 100, 160, 250, 350 and 500$\,\mu$m photometry from the GOODS-{\it Herschel} survey (Elbaz et al.~2011). 
We fit all the data to the empirical spectral energy distribution (SED) templates from Kirkpatrick et al.~(2012) which were derived from similarly high redshift star-forming galaxies. 
The {\it Herschel} photometry of these sources can be found in Kirkpatrick et al.~(2012) and we list the calculated IR luminosities in Table~\ref{tab:sample}. The error on the $L_{\rm{IR}}$ is conservatively calculated to be 40\% which includes the uncertainty due to the scatter in the SED shapes and the photometric uncertainties (Kirkpatrick et al.~2012). 
We find that these 70$\,\mu$m-selected galaxies have IR luminosities of 0.7--2.6$\times10^{12}L_{\odot}$; all but one are formally classified as ULIRGs. 


\subsection{IRAM PdBI observations}
\label{sec:obs}

Observations of three 70$\,\mu$m-selected sources were taken with the PdBI over several days in August and September 2009 with 5 antennas\footnote{Two hours of data for GN70.104 was taken with only 4 antennas.} in D configuration.  On-source integration times were 7.4, 9.9 and 6.2 hours for GN70.8, GN70.104 and GN70.14, respectively. The 3$\,$mm receiver was tuned to the expected frequency of CO(2-1) based on the optical spectroscopic redshifts from Wirth et al.~(2004) and Barger et al.~(2008).
All targets were observed in good observing conditions with typical system temperatures of $\sim115\,$K. 

Data were reduced and calibrated using the IRAM GILDAS package\footnote{http://www.iram.fr/IRAMFR/GILDAS}. 
We used 1344+661 and 1418+546 for the phase calibration and 3c454.3 and 3c345 for the bandpass calibration.  
The flux calibration was done using CRL618 and MWC349; the uncertainty in the flux calibration at 3mm with IRAM PdBI is typically 10\%.

Two of three sources are detected in our observations and both detections are unresolved. The signal-to-noise ratio of the integrated line intensity is 4.4 for GN70.8 and 11.4 for GN70.104. These CO detections are bolstered by the coincidence with the optical galaxy, spatially and in velocity space given the optical spectroscopic redshift (see Section \ref{sec:co} for more details). The final beams in the maps of the two detected sources are 5.33 by 4.58 arcseconds (PA=43.4 degrees) for GN70.8 and 5.63 by 4.86 arcseconds (PA=71.42 degrees) for GN70.104. The measured CO properties of the two detected sources are given in Table~\ref{tab:co}. Figs.~\ref{fig:co1} and~\ref{fig:co2} show the CO spectra and the contours of CO line emission on the {\it HST} images (Giavalisco et al.~2004).

\subsection{{\it Spitzer} IRS observations}
\label{sec:irs}

\begin{figure*}
\begin{center}
\includegraphics[width=5.0in,angle=0]{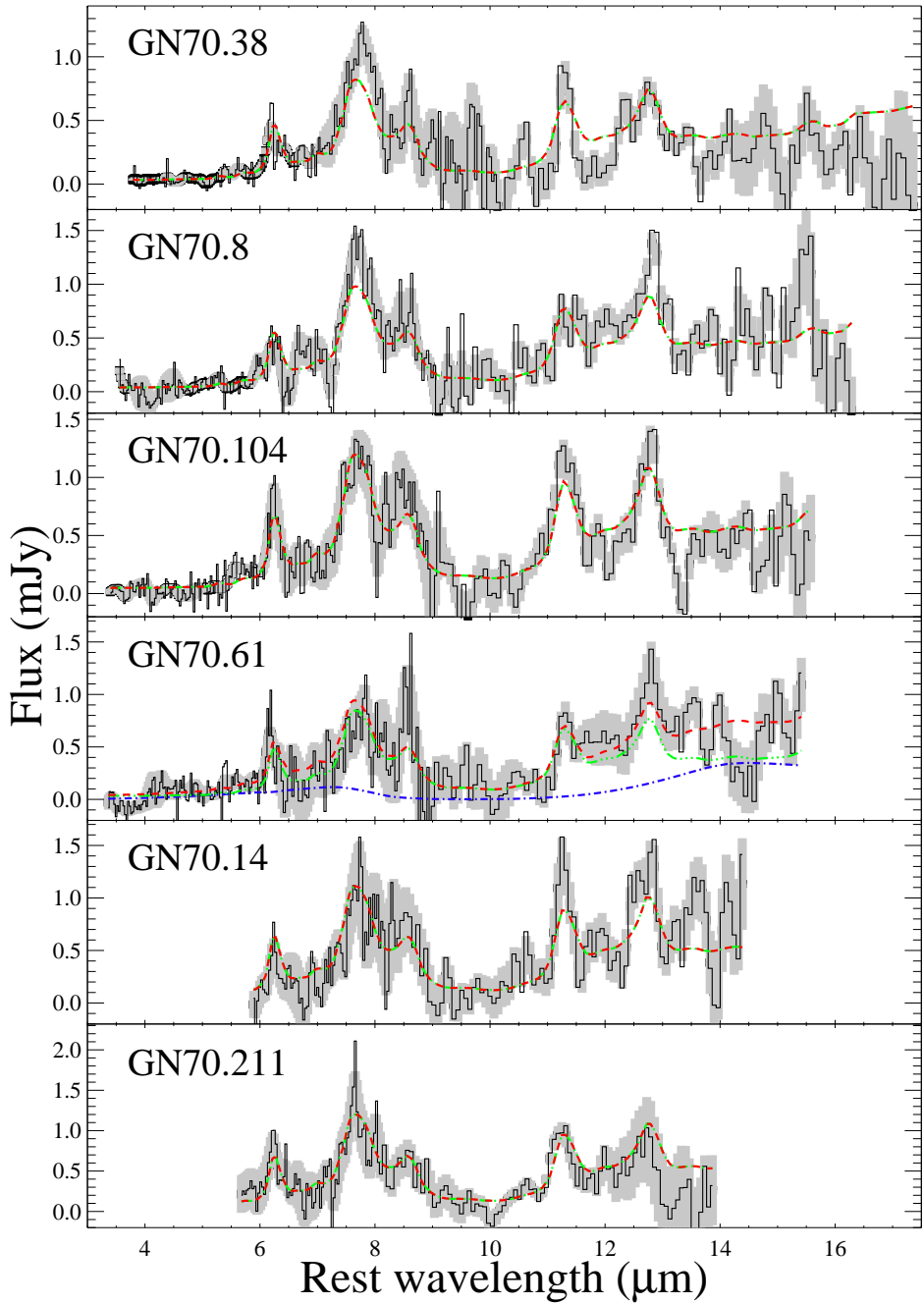}
\vspace{0.1in}
\caption{New {\it Spitzer} IRS spectra of 70$\,\mu$m-selected ULIRGs. The black solid curves and associated grey regions indicate the raw (unsmoothed) mid-IR spectra and 1$\sigma$ uncertainties. The results of our spectral decomposition are shown as the other curves: the green dash-dot-dot is the starburst component, the blue dash-dot is the AGN continuum component with extinction and the red dashed line is the combined best-fit model. Only one of these sources (GN70.61) has a best-fit model which includes a significant AGN component.
}
\label{fig:pah}
\end{center}
\end{figure*}

We obtained IRS observations of six 70$\,\mu$m-selected galaxies through {\it Spitzer} program DDT 288 in May and June 2007. 
Observations were taken at low resolution in spectral staring mode with the target at two different nod positions within the slit to facilitate sky subtraction. 
We chose the IRS orders that would cover the PAH features from 6--13$\,\mu$m and integration times were estimated from the 24$\,\mu$m flux to reach the same signal-to-noise ratio in the resulting spectra.
Table~\ref{tab:pah} summarizes the specifics of the IRS observations for each source. 

The IRS data were reduced following the procedure outlined in Pope et al.~(2008a). To summarize, we began with the two-dimensional basic calibrated data from the {\it Spitzer} pipeline and began by cleaning the rogue pixels and removing the latent build-up on the arrays. Next, we performed sky subtraction by creating a normalized `supersky' from all observations in the same astronomical observation request (AOR). Detailed tests have shown that this method results in the lowest residual sky noise in the final spectrum (Pope et al.~2008a). Individual sky-subtracted two-dimensional data files were co-added and then checked again for rogue pixels. 

The Spitzer IRS Custom Extraction (SPICE) software was used to extract the one-dimensional spectrum at each nod position for each source in optimal extraction mode. The two one-dimensional spectra at each nod position are then coadded to create a final reduced spectrum for each source. 
For each target, we also extract a residual sky spectrum which is used to quantify the uncertainty in the final target spectrum. 
The reduced one-dimensional spectra are shown in Fig.~\ref{fig:pah}. 

In order to quantify the contribution from AGN and star formation activity to the mid-IR emission, we decompose the IRS spectra using the method outlined in Pope et al.~(2008a). A model consisting of 3 components is fit to the spectrum: a PAH template (either M82 or the starburst composite from Brandl et al.~2006), a continuum component with slope as a free parameter, and an additional extinction component from the Draine (2003) models which includes the prominent 9.7$\,\mu$m silicate absorption feature. The resulting fits are also shown in Fig.~\ref{fig:pah} and the measured quantities are listed in Table \ref{tab:pah}. 
When the best-fit model does not include a continuum component, we calculate an upper limit by forcing a continuum to fit between the PAH features. 
This spectral decomposition provides an upper limit to the AGN fraction in the mid-IR since we assume all the continuum emission is coming from the AGN while there will be some continuum emission from star formation (e.g.,~Tran et al.~2001). 

An alternative way to quantify the amount of star formation activity in the mid-IR is through the strengths of the individual PAH lines. We measure the luminosity of the 6.2$\,\mu$m PAH line as it is least affected by the silicate absorption and blending with other PAH lines. We measure the strength of the line above a spline continuum (see Brandl et al.~2006) since this is the most robust for low signal-to-noise ratio spectra. Individual PAH luminosities can vary considerably depending on how the measurements are made (see discussion in Smith et al.~2007), but we note that, in the absence of very strong silicate absorption, relative comparisons will be robust if the lines are measured in the same way for all sources.

\section{Literature data}

In order to interpret the results for our sample of $z\sim1$ ULIRGs, we compile a list of comparable observations for other galaxies locally and at high redshift from the literature. Specifically, we search for all other sources which have both CO and PAH detections. We limit ourselves to sources which are not obviously lensed since differential lensing could complicate the comparison of CO and PAH emission.

\subsection{High redshift ULIRGs}
\label{sec:hizsample}

At $z>1$ we find 12 ULIRGs with detectable 6.2$\,\mu$m PAH emission in their IRS spectrum and with CO detections including the new observations presented in this paper. These sources are all located in GOODS-N which allows us to calculate $L_{\rm{IR}}$ using the same data and SED fitting that we do for the 70$\,\mu$m-selected galaxies. This heterogeneous sample includes SMGs, BzKs and 70$\,\mu$m-selected ULIRGs. This is not a complete sample of high redshift galaxies but it does span several main sub-populations of high redshift ULIRGs which have been studied extensively in the literature. The high redshift literature data and associated references are listed in Table~\ref{tab:hiz} including a column which gives the galaxy type and identifies any overlap between the selections (which is simply limited by the depth of the data). 

In order to make a fair comparison, we re-reduced all the IRS data ourselves and made consistent PAH line measurements. The total IR luminosity measurements are made in the same way as for the 70$\,\mu$m-selected galaxies by fitting all available {\it Spitzer} and {\it Herschel} data to the empirical Kirkpatrick et al.~(2012) templates (see Section \ref{sec:samp}). 

For the CO data (not publicly available) we use the CO fluxes in the literature and recalculated all luminosities under the same cosmological model. 
When multiple CO transitions are available for the same source we list the $L^{\prime}_{\rm{CO}}$ for the lowest J transition since ultimately we will convert all CO luminosities down to the (1-0) transition for comparisons. For the two BzK galaxies and two of the SMGs we use direct measurements of the (1-0) transition. For the remaining sources, we use the following conversions from Frayer et al.~(2011): $r_{21}=L^{\prime}_{\rm{CO(2-1)}}/L^{\prime}_{\rm{CO(1-0)}}=0.8\pm0.1$, $r_{31}=L^{\prime}_{\rm{CO(3-2)}}/L^{\prime}_{\rm{CO(1-0)}}=0.6\pm0.1$, and $r_{41}=L^{\prime}_{\rm{CO(4-3)}}/L^{\prime}_{\rm{CO(1-0)}}=0.4\pm0.1$. 
These luminosity, or brightness temperature ratios have been derived for submm luminous star-forming galaxies and should also be appropriate for 70$\,\mu$m-selected ULIRGs. Uncertainties that arise from these assumed brightness temperature ratios are small enough that they do not change any of the trends or results presented in this paper. 

We also include a larger sample of high redshift galaxies with {\it Spitzer} IRS spectra in some of our analysis. For this we use the sample of 24$\,\mu$m-selected high redshift (U)LIRGs from Kirkpatrick et al.~(2012). This sample consists of 151 sources in GOODS-N and ECDFS with deep IRS spectra and deep {\it Herschel} data for calculating $L_{\rm{IR}}$ (GOODS-{\it Herschel}, Elbaz et al.~2011). 
We cull this sample by taking sources with $z=1$--3, a detection of the 6.2$\,\mu$m PAH line, an AGN fraction $<50\%$ (since we want to be certain $L_{\rm{IR}}$ is dominated by star formation), and sufficient data in the far-IR to get a good measure of $L_{\rm{IR}}$. 
To be free from any systematic biases, we calculate the PAH and total IR luminosities of this sample in the exact same way as our main 70$\,\mu$m-selected targets and the other high redshift galaxies described above. 
Unfortunately we do not have CO measurements for the full Kirkpatrick et al.~(2012) sample.

\subsection{Local ULIRGs}
\label{sec:localsample}

We compare the high redshift ULIRGs to local ULIRGs with both PAH and CO detections. 
Armus et al.~(2007) presented IRS spectra for a sample of BGS ULIRGs, all of which have CO detections. Since a large fraction of these ULIRGs are AGN dominated in the mid-IR with weak or no PAH emission, we supplement the local sample with some strong PAH ULIRGs from the larger IRS local ULIRG sample of Desai et al.~(2007). In total we include 14 local ULIRGs with CO and PAH detections; 7 from Armus et al.~(2007) and 7 from Desai et al.~(2007). 
We apply our same mid-IR spectral decomposition to this sample of local ULIRGs to ensure that PAH luminosity measurements are performed in a consistent way to facilitate comparisons with the high redshift ULIRGs. 
We find CO(1-0) measurements for all local ULIRGs from Sanders et al.~(1991), Solomon et al.~(1997), Gao \& Solomon~(2004) and Chung et al.~(2009), and re-calculate the CO luminosities under our assumed cosmology. Total IR luminosities for the local ULIRGs are also provided in the literature although uncertainties are not given. A recent paper by U et al.~(2012) fit the SEDs for a subset of northern local ULIRGs using all the recent {\it Spitzer} data and found very similar values of $L_{\rm{IR}}$ to previous IRAS estimates and quantify the uncertainty in $L_{\rm{IR}}$ to be small, on order of 5\% for most sources. 
In total we have 14 local ULIRGs with both CO and PAH measurements (Table \ref{tab:local}), a comparable size to the high redshift sample. 

At lower IR luminosities, we use nearby sources from the 5MUSES IRS spectroscopic survey for comparison. We obtain the IR and PAH luminosities from Wu et al.~(2010). Their PAH luminosities are measured using PAHFIT which is known to produce $L_{\rm{PAH,6.2}}$ values a factor of 1.7 higher than the spline-based line decomposition which we have used for the high redshift sources (Smith et al.~2007). 
We globally correct their $L_{\rm{PAH,6.2}}$ values for this factor of 1.7 to provide a good comparison to our high redshift measures. The median redshift for the 5MUSES sources is 0.08. Unfortunately we do not have CO measurements for the 5MUSES sample.

\section{Results}

\begin{figure*}
\begin{center}
\includegraphics[width=2.0in,angle=0]{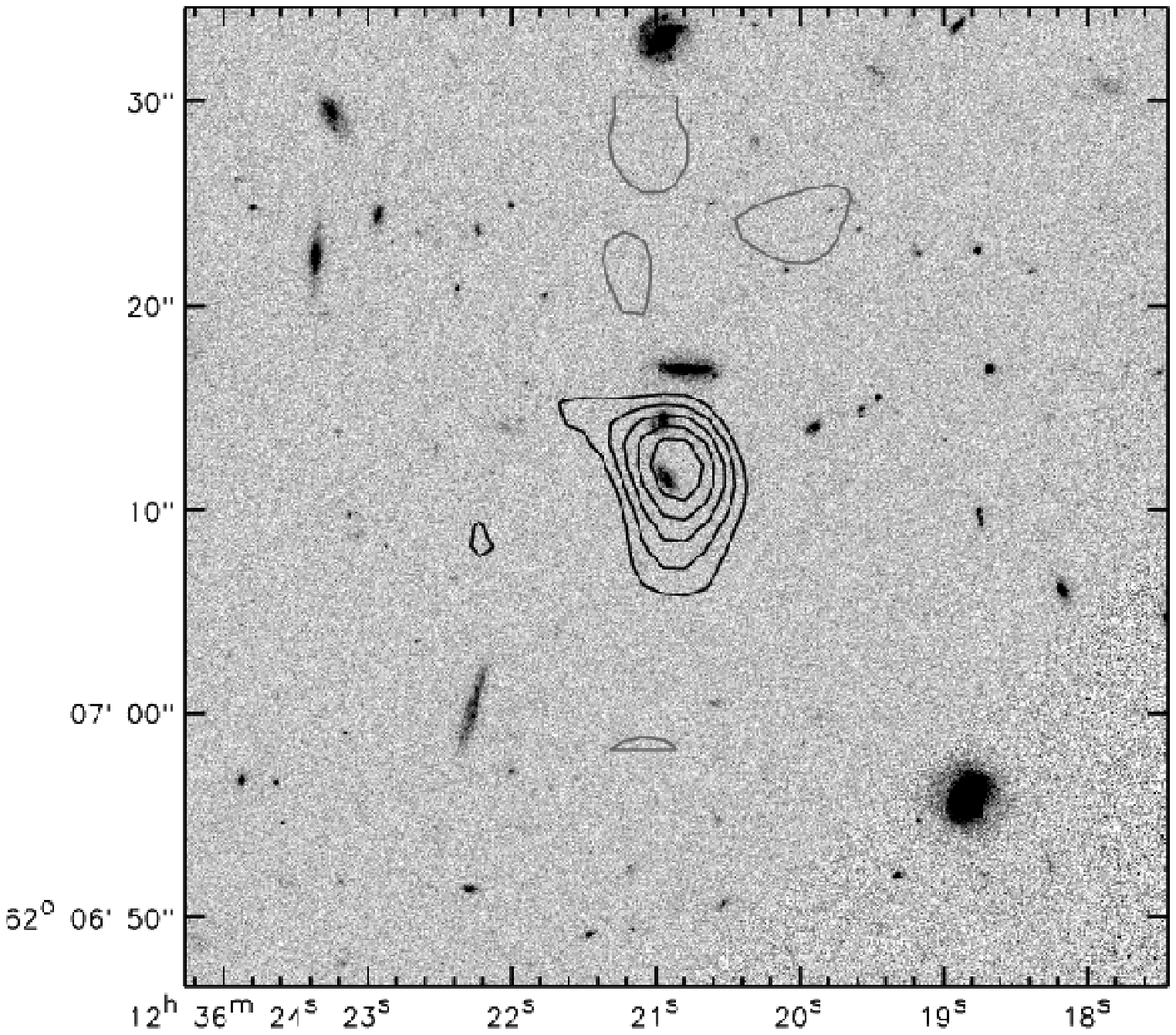}
\includegraphics[width=2.0in,angle=0]{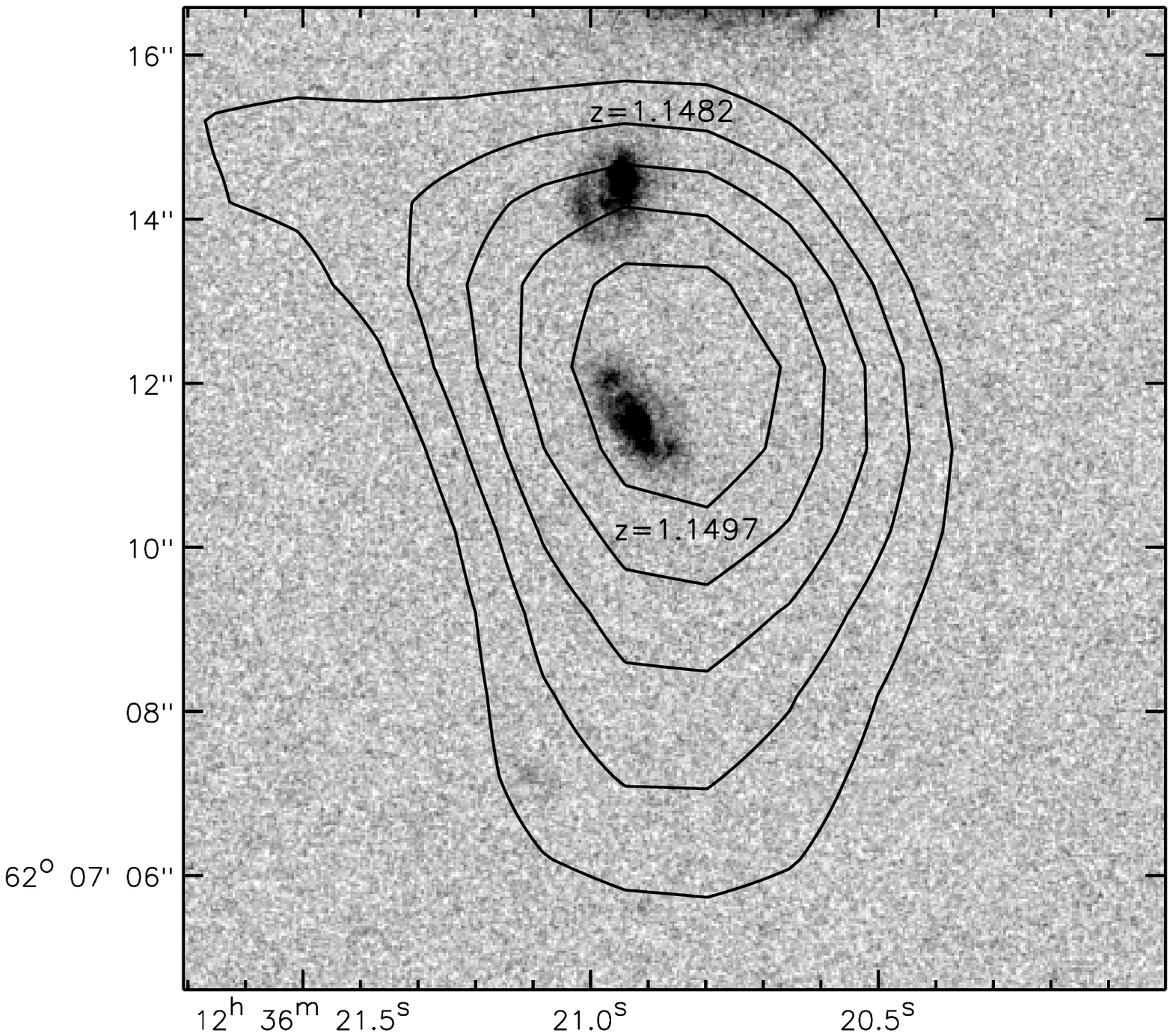}
\includegraphics[width=2.0in,angle=0]{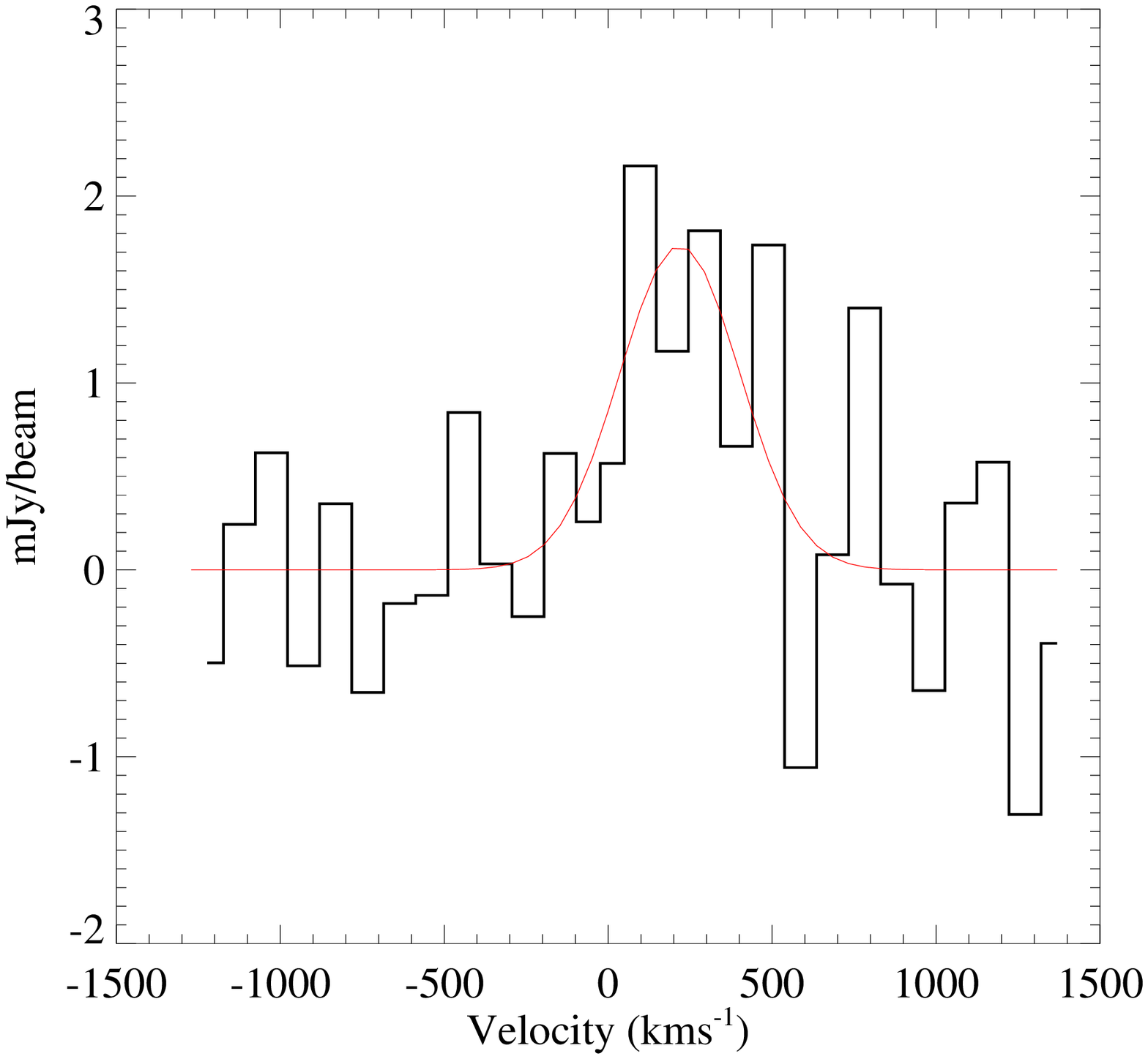}
\caption{CO (2-1) observations of GN70.8. The left and middle panels shows the contours of CO line emission (2, 3, 4, 5, 6$\sigma$ averaged over 600$\,$km/s to maximize the SNR; positive contours in black and negative contours in light grey) on the {\it HST} ACS $i$-band optical image while the right panel shows the CO spectrum. 
The close pair of optical galaxies is clearly indicated by the spectroscopic redshifts in the middle panel. Our resolution is only 5 arcseconds FWHM; therefore, we are unable to determine if the CO emission is spatially offset or extended between the two optical sources. 
}
\label{fig:co1}
\end{center}
\end{figure*}

\begin{figure*}
\begin{center}
\includegraphics[width=2.0in,angle=0]{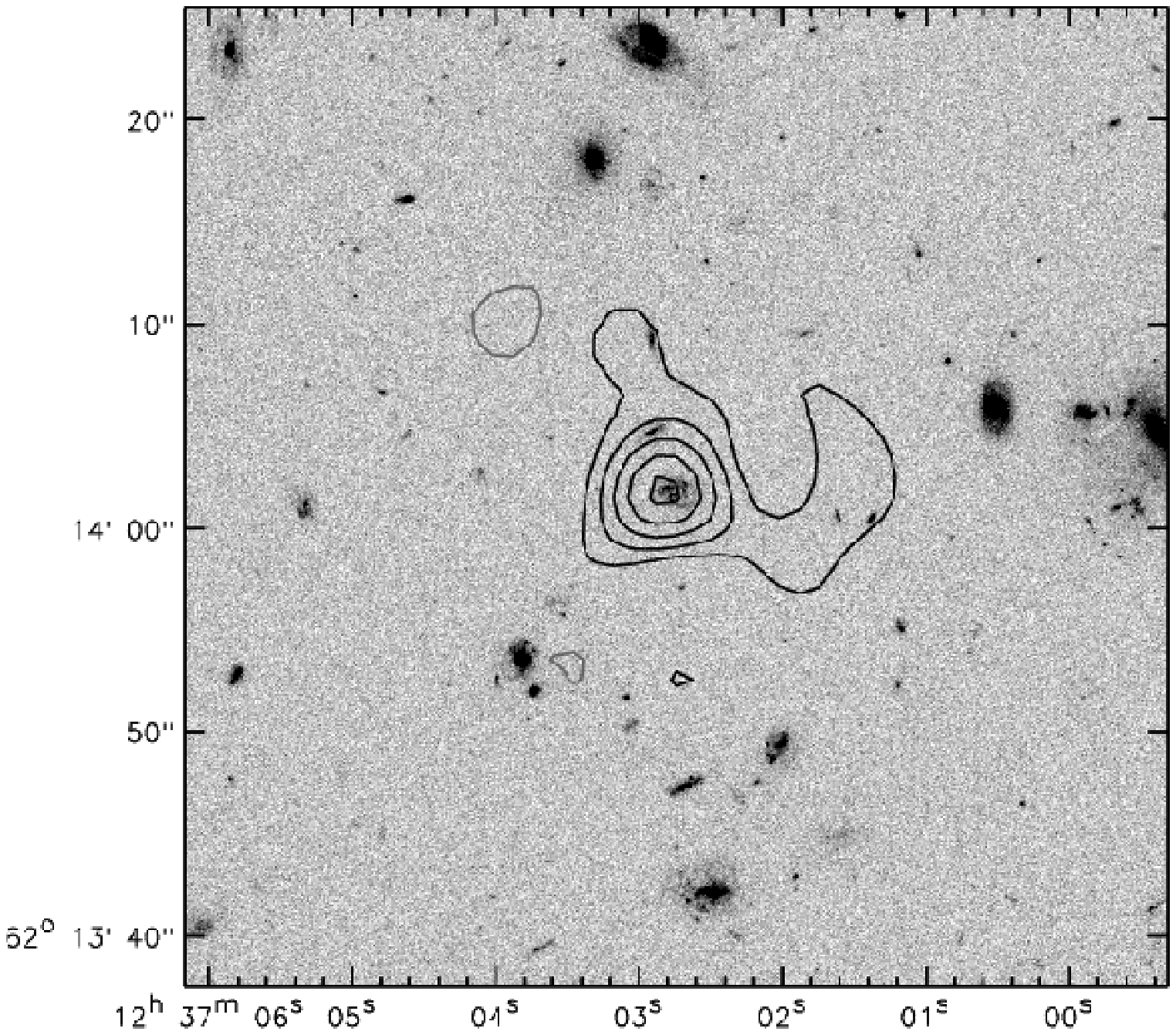}
\includegraphics[width=2.0in,angle=0]{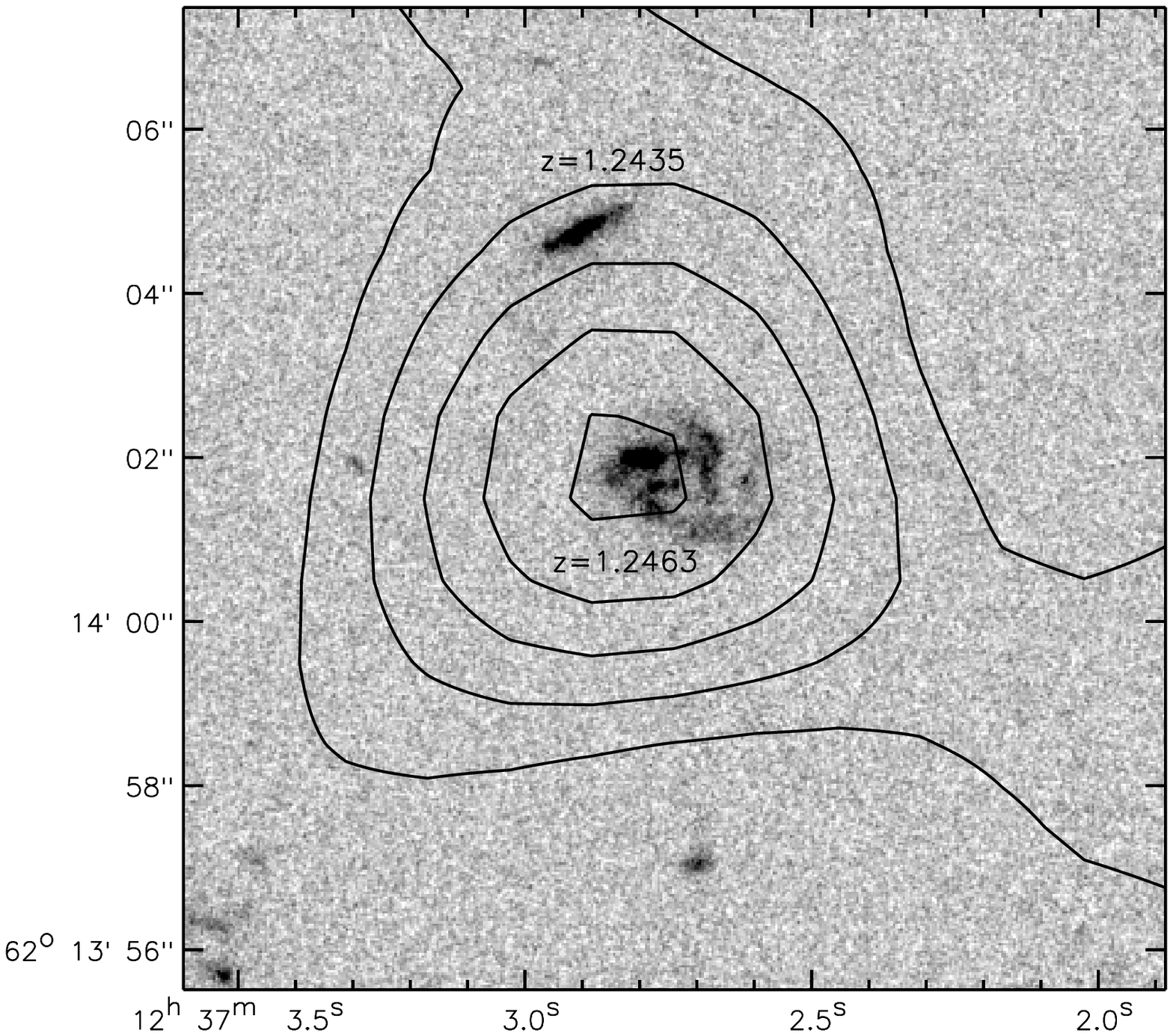}
\includegraphics[width=2.0in,angle=0]{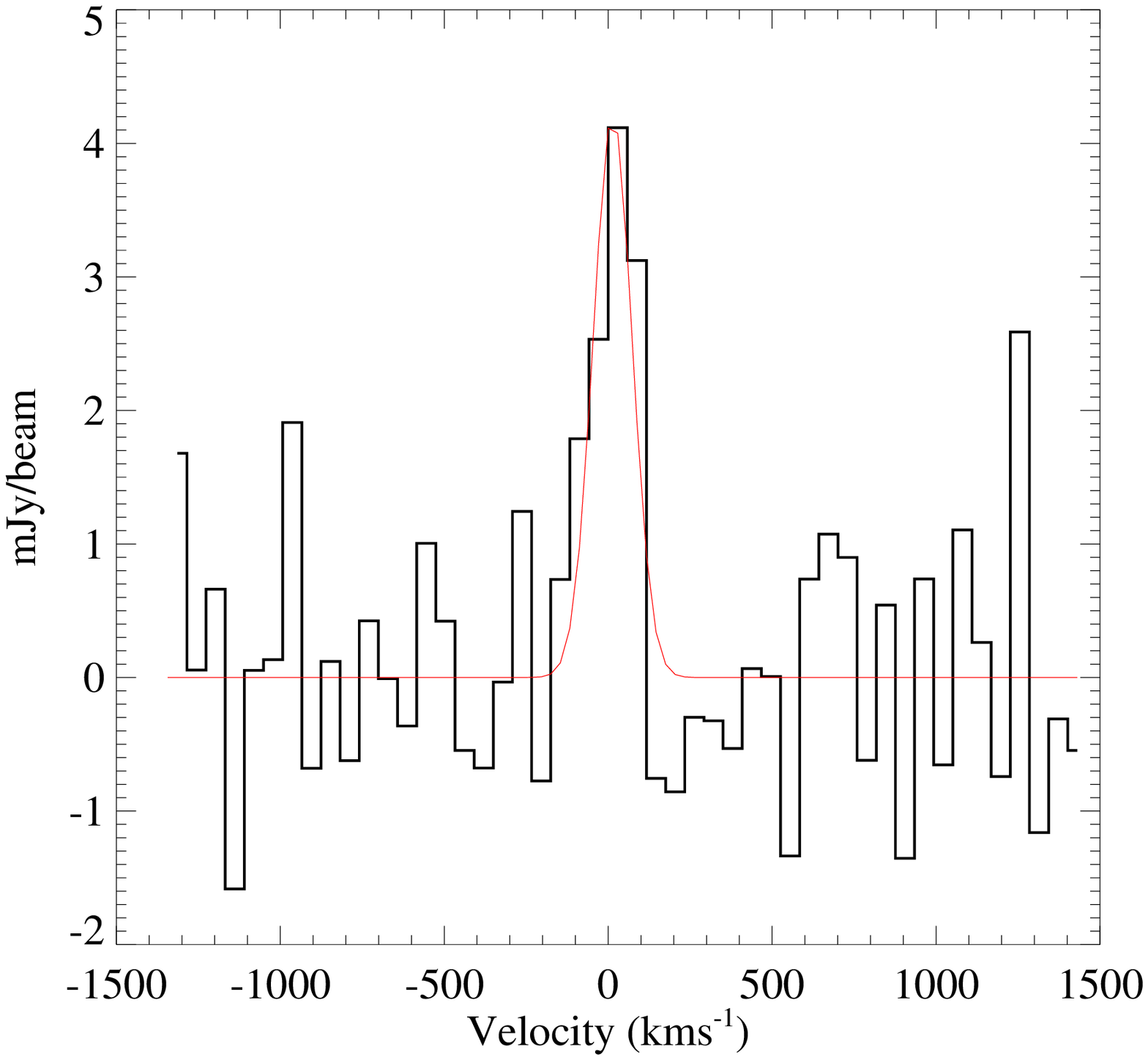}
\caption{CO (2-1) observations of GN70.104. The left and middle panels shows the contours of CO line emission (2, 4, 6, 8, 10$\sigma$ averaged over 150$\,$km/s to maximize the SNR; positive contours in black and negative contours in light grey) on the {\it HST} ACS $i$-band optical image while the right panel shows the CO spectrum. 
The close pair of optical galaxies is clearly indicated by the spectroscopic redshifts in the middle panel. 
}
\label{fig:co2}
\end{center}
\end{figure*}

\subsection{CO molecular gas properties}
\label{sec:co}
We detected CO line emission from two 70$\,\mu$m-selected galaxies, both of which are found to be close pair systems: two close optical counterparts with consistent spectroscopic redshifts. Since the resolution of the compact configuration CO observations is $\sim5\,$arcseconds, the CO emission within the pairs is not spatially resolved.

GN70.8 has two optical counterparts separated by 2.8 arcseconds (23$\,$kpc) and 440$\,$km/s (derived from the optical spectroscopic redshifts). The peak of the MIPS 24$\,\mu$m emission lies between the two optical galaxies. 
From the background {\it HST} image in Fig.~\ref{fig:co1}, these two galaxies appear to be interacting with evidence of asymmetries and possible tidal features. 
This is further supported by the peak of the CO emission in velocity space which is shifted between the two redshifts, 210$\,$km/s from the lower redshift source. 
The CO emission could be spatially centered and/or extended between this close pair, although the spatial resolution is not sufficient to resolve or definitively localize the emission. 

GN70.104 has two optical counterparts separated by 3.2 arcseconds (27$\,$kpc) and 840$\,$km/s (derived from the optical spectroscopic redshifts). Again the MIPS 24$\,\mu$m emission is unresolved between the two optical galaxies. 
Fig.~\ref{fig:co2} shows that CO emission is more consistent with optical galaxy to the south both spatially and in velocity space, although with our resolution and sensitivity we are unable rule out some CO emission from the other optical galaxy.

\begin{figure*}
\begin{center}
\includegraphics[width=5.0in,angle=0]{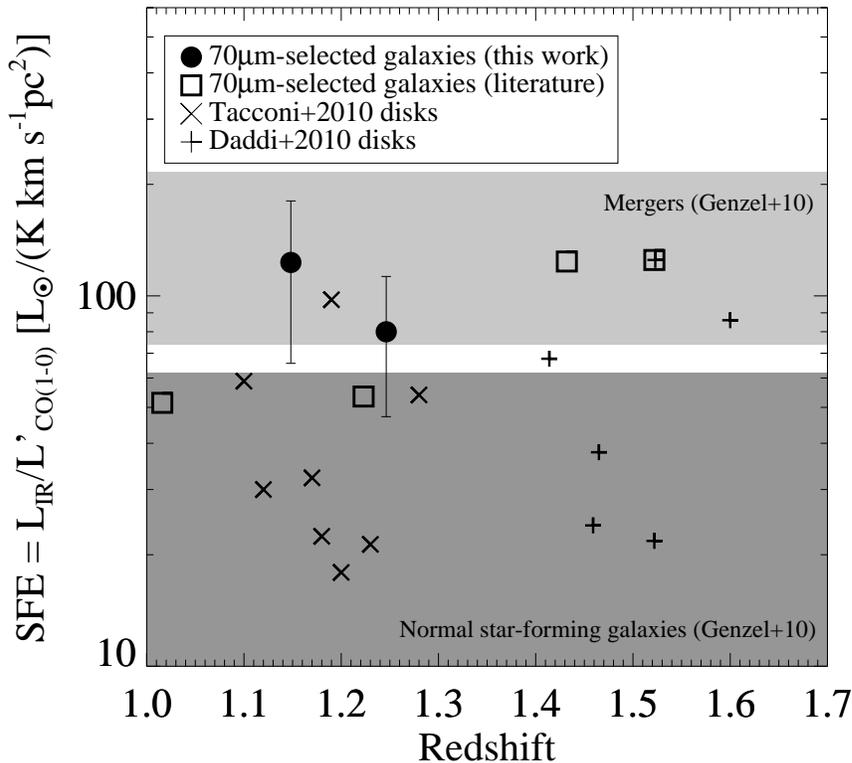}
\caption{Star formation efficiency in galaxies at $z\sim1$--2. We plot our new observations of 70$\,\mu$m-selected galaxies compared to other high redshift galaxies in the literature. 
The shaded regions show the SFE (average value $\pm$ 1 standard deviation from Genzel et al.~2010) for local and high redshift normal star-forming galaxies (dark grey) and mergers (local ULIRGs and high redshift SMGs, light grey). 
Our new measurements of 70$\,\mu$m-selected ULIRGs which are close pairs (filled circles) have similar SFEs to mergers locally and at $z\sim2$ and lie at relatively unexplored intermediate redshifts. 
}
\label{fig:sfe}
\end{center}
\end{figure*}

GN70.14 also has two optical counterparts separated by 600$\,$km/s in velocity space and 1 arcsecond (8.5$\,$kpc) spatially. This source was undetected in our CO measurements but was also observed by Casey et al.~(2011) who report a detection of CO (2-1) with an intensity of $1.8\pm0.5\,$Jy$\,$km/s. The upper limit from the non-detection in our data is consistent with their value given the uncertainties. We include this source in the literature sample in Table ~\ref{tab:hiz}.

We stress that our targets were not pre-selected to be close pairs, we simply selected them as 70$\,\mu$m detected galaxies with an available spectroscopic redshift around $z\sim1$. The fact that all three that we observed in CO are close pairs is not terribly surprising. Kartaltepe et al.~(2010) found that at least 60\% of 70$\,\mu$m-selected ULIRGs at $z=1$--1.5 are major or minor mergers based on an optical morphological study. Furthermore, they found that roughly a quarter of galaxies with similar luminosity to our 70$\,\mu$m sources are in early merger stages before the two nuclei have coalesced. From theoretical modeling, Hayward et al.~(2013) show that a significant fraction of the number counts of the most IR luminous galaxies may come from two well separated galaxies both falling in the large far-IR/submm beam; the combined flux of close pairs may be more likely to be detected in the far-IR/submm even if an interaction is not inducing a burst. 

In the local merging system II Zw 096, Inami et al.~(2010) found that almost all the IR luminosity originates from a compact dusty source not associated with either of the nuclei of the merging galaxies. High spatial resolution CO observations are needed to determine the position and spatial extent of the molecular gas in these close pairs to determine how the interaction might be affecting the distribution of molecular gas. The signal-to-noise in our spectra is not high enough to detect multiple peaks in the line profiles and the relatively narrow linewidths observed are consistent with those observed in quasar host galaxies and BzK-selected star-forming galaxies (Carilli and Wang 2006; Coppin et al.~2008).

Another indicator of merger-induced star formation activity is the SFE, defined as the ratio of the total IR luminosity (which traces the star formation rate) to the CO luminosity (which traces the total molecular gas mass). The inverse of this ratio is related to the gas depletion timescale. 
The dense gas fraction is observed to be much higher in a merger-induced burst than in isolated star forming galaxies and the former have higher SFEs (Gao \& Solomon~2004). 
In Table ~\ref{tab:co} we list the observed CO(2-1) luminosity as well as the estimated CO(1-0) luminosity.  High redshift normal disk galaxies and SMGs both show slightly sub-thermal excitation with $r_{21}\sim0.8\pm0.1$ (Aravena et al.~2010; Frayer et al.~2011). 
In Fig.~\ref{fig:sfe}, we plot the SFE ($=L_{\rm{IR}}$/$L^{\prime}_{\rm{CO(1-0)}}$) as a function of redshift over the intermediate redshift range spanned by our targets. 
The grey shaded regions on the figure show the range of SFEs (average value $\pm$ 1 standard deviation) observed for normal star forming galaxies (dark grey) and mergers (light grey) from Genzel et al.~(2010). The normal star-forming galaxies consist of high redshift main sequence galaxies observed in CO as well as local normal star-forming galaxies, while the mergers include local ULIRGs and high redshift SMGs. 
We also overplot individual observations of disk galaxies (most of which are not 70$\,\mu$m detected) from Tacconi et al.~(2010) and Daddi et al.~(2010a), and other 70$\,\mu$m-selected galaxies that fall within this redshift range. 
The Daddi et al.~(2010a) disks are plotted using the updated $L_{\rm{IR}}$ values based on {\it Herschel} data (Magdis et al.~2012). 
Our new CO observations of the 70$\,\mu$m close pairs have high SFEs at their redshifts and are consistent with the range of SFE expected for mergers. 
Our two 70$\,\mu$m-selected close pairs have very similar stellar masses (few$\times10^{11}M_{\odot}$) to the Tacconi et al.~(2010) disks at the same redshift and so stellar mass is not causing the separation. 
It is interesting to note that some galaxies classified as disks end up with high SFEs consistent with mergers while a few galaxies which have disturbed morphologies (e.g.~GN26, open square plotted at $z=1.22$, see discussion in Frayer et al.~2008) have lower SFEs similar to the normal star-forming galaxies. 
It is important to remember that these SFE regions represent the average $\pm$ 1 standard deviation for each category and larger samples of high redshift galaxies may show a more continuous distribution of SFE spanning both regions. We must be cautious in placing small subsamples of galaxies within this picture.

The high SFEs for our new targets are intriguing since they are very well separated pairs (Figs.~\ref{fig:co1} and \ref{fig:co2}). 
Many merger simulations show that the maximum boost in the SFR occurs after coalescence (e.g.,~Narayanan et al.~2010), although these simulations start with a constant gas fraction and do not allow inflow of gas to the system during the merger. 
However, Mihos \& Hernquist (1996) show that if the merging galaxies are simple disks without bulges, as many high redshift galaxies are, that the SFR during the first-pass and subsequent maximum separation is even higher than at coalescence (see also Di Matteo et al.~2007). 
Our results show enhanced SFR (relative to the CO luminosity) in these close pairs, suggesting that the interaction does have an effect on the SFR even during these early merger stages. The ratio of $L_{\rm{IR}}$/$L^{\prime}_{\rm{CO}}$ may also be affected by other factors (such as gas fractions, excitation, environment, etc.) which are not fully taken into account here. 
While it is tempting to use this simple ratio to separate merger-induced starbursts from simple disks, our new observations reinforce the idea that we need need more studies of early merger stages, when galaxies may be seemingly isolated, to better quantify observationally how much of the star formation is triggered by the interaction.

In Table \ref{tab:co}, we also list the total molecular gas mass ($M_{\rm{H_{2}}}$). The calculation of the total molecular gas mass from the observed CO luminosity requires an assumption of the CO-H$_{2}$ conversion factor ($\alpha_{\rm{CO}}$). 
While there are some observational constraints on $\alpha_{\rm{CO}}$ at high redshift for normal star forming galaxies and high redshift mergers (Daddi et al.~2010a; Tacconi et al.~2008; Magdis et al.~2012), theoretical models find that even within these regimes there is a wide range of acceptable values (see e.g.,~Narayanan et al.~2011, 2012). 
Furthermore, our targets are close pairs which are likely to be early stage mergers; however, it is not clear if the interaction has progressed enough to cause enhanced star formation or to affect the value of $\alpha_{\rm{CO}}$. For these reasons, we quote a range of $M_{\rm{H_{2}}}$ for each source assuming the range of conversions spanned by normal star forming galaxies and mergers.

\subsection{PAH dust properties}

The reduced 1D mid-IR spectra are shown in Fig.~\ref{fig:pah} along with the results of the spectral decomposition to quantify the relative amounts of PAH (green curve) and continuum emission with extinction (blue curve) to the best-fit model (red curve). The PAH component dominates in all sources indicating that the majority of the mid-IR luminosity is coming from star formation since PAHs are suppressed or even destroyed around AGN (e.g.,~Genzel et al.~1998; Weedman et al.~2005; Armus et al.~2007; Desai et al.~2007). GN70.61 has the largest contribution from the continuum component assumed to come from the AGN; Table \ref{tab:pah} quantifies this contribution as 24\% of the mid-IR luminosity. 

We stress that these sources were selected for mid-IR spectroscopy solely based on having a 70$\,\mu$m detection, an optical spectroscopic redshift and a 24$\,\mu$m flux bright enough to be observed with IRS in a reasonable amount of time ($S_{24}>300\,\mu$Jy). The low AGN fraction is likely affected by this selection. Given that we are looking in a small field (GOODS-N is 160 square arcminutes) with relatively deep 70$\,\mu$m observations, our selection picks out sources near the 70$\,\mu$m detection limit ($S_{70}=2$--7$\,$mJy), with higher $S_{70}$/$S_{24}$ ratios since sources at $z\sim1$ with $S_{24}\gtrsim1\,$mJy are rare. 
Brandl et al.~(2006) find that local starburst galaxies with weak or no AGN all have rest-frame colors of $S_{30}$/$S_{15}>5$, which is roughly equivalent to $S_{70}$/$S_{24}>5$ at $z\sim1$. 
The average $S_{70}$/$S_{24}$ for our sample is 12 (standard deviation is 5) indicating that the majority of our sample is above the threshold found in Brandl et al.~(2006) for local starburst galaxies. Most AGN dominated sources at $z\sim1$ have lower ratios of $S_{70}/S_{24}\sim2$--4 (Mullaney et al.~2010; Kirkpatrick et al.~2013); none of our sources have flux ratio this low. 

The mid-IR AGN fraction is also a strong function of the mid-IR and IR luminosity (e.g.,~Tran et al.~2001; Desai et al.~2007; Dey et al.~2008; Donley et al.~2008). Our 70 microns targets have $L_{\rm{IR}}\simeq1$--$3\times10^{12}\,L_{\odot}$ (Table \ref{tab:sample}); at these IR luminosities at $z\sim1$ we don't expect a large fraction of galaxies to be dominated by an AGN. 
Farrah et al.~(2009) present IRS spectra of a brighter flux-selected sample of 70$\,\mu$m sources from a much wider, shallower survey ($S_{70}>19\,$mJy); more than half of their sample at $z>0.8$ have measured 6.2$\,\mu$m PAH equivalent widths less than 0.2$\,\mu$m indicating that they are dominated by AGN in the mid-IR (Armus et al.~2007; Sajina et al.~2007; Pope et al.~2008a).

Besides confirming that star formation is dominating the energetics of these 70$\,\mu$m-selected galaxies in the mid-IR, we can also measure the luminosity of PAH features to further quantify the emission from small dust grains which is shown to trace the SFR (e.g.,~Brandl et al.~2006; Pope et al.~2008a). The measured $6.2\,\mu$m PAH luminosities which we use in this paper are listed in Table~\ref{tab:pah}. We do not correct the PAH luminosities for extinction since the best-fit model for each source from our spectral decomposition requires no additional extinction. We do not use the 7.7 and 8.6$\,\mu$m luminosities as these features are strongly affected by blending which can bias their measurements. 
In Section \ref{sec:dustgas}, we explore how these PAH luminosities are related to the IR and CO luminosities for these $z\sim1$ ULIRGs as compared to local ULIRGs and $z\sim2$ SMGs allowing us to probe the relationship between gas and dust in the interstellar medium at high redshift.

\subsection{Normal versus starburst galaxies}

The correlation observed between the star formation rate and the stellar mass in high redshift galaxies has been interpreted to indicate that most stellar growth in galaxies is occurring continuously and not through violent, short-lived starburst events (e.g.,~Noeske et al.~2007; Elbaz et al.~2011). The track followed by most galaxies in the SFR-$M_{\ast}$ plane has been called the `main sequence', and galaxies which lie significantly above the main sequence (higher SFR for their $M_{\ast}$) have been grouped as `starbursts'. The general idea put forward is that main-sequence galaxies are fueled by continuous gas accretion whereas starbursts may be undergoing additional external triggers such as mergers or interactions. 
While this idea is difficult to test for all galaxies at high redshift, it does hold in the local Universe where all SDSS galaxies lie on the SFR-$M_{\ast}$ relation and galaxies above the main sequence are nearly always mergers (e.g.~Salim et al.~2007; Peng et al.~2010). Recently, Elbaz et al.~(2011) noticed that main sequence galaxies locally and at high redshift also follow a similar correlation in $L_{\rm{IR}}$ vs $L_{\rm{8\mu m}}$, while starburst galaxies have higher $L_{\rm{IR}}$ for their $L_{\rm{8\mu m}}$ relative to this relation. 

It is interesting to see where our galaxies lie in the SFR-$M_{\ast}$ and $L_{\rm{IR}}$-$L_{\rm{8\mu m}}$ planes relative to known mergers and isolated galaxies.  
In Table \ref{tab:sample}, we list the specific star formation rate (sSFR=SFR/$M_{\ast}$) and the ratio of $L_{\rm{IR}}/L_{\rm{8\mu m}}$ (=IR8) for all six of the 70$\,\mu$m-selected galaxies for which we obtained deep IRS spectra showing strong PAH emission. SFRs are estimated from $L_{\rm{IR}}$ assuming the standard Kennicutt (1998) conversion. 
Stellar masses are estimated using the same method as Elbaz et al.~(2011) so we can directly compare the sSFRs of our galaxies to the Elbaz main sequence criteria. Stellar masses are derived from fitting the U-band to IRAC 4.5$\,\mu$m data using Z-PEG assuming a Salpeter IMF (Salpeter 1955) and allowing for nine different star formation histories (Le Borgne \& Rocca-Volmerange 2002). 
In cases where there is a pair of optical galaxies within the 24$\,\mu$m beam, we calculate the total stellar mass and SFR of the whole system. If the star formation is only associated with one of the galaxies in the pair then the sSFR will be underpredicted. 
In Table \ref{tab:sample}, we indicate whether our 70$\,\mu$m-selected galaxies are classified as main sequence (MS) or starbursts (SB) based on the relations derived in Elbaz et al.~(2011, Equations 2 and 14) for galaxies at similar redshifts. 
We notice that for most galaxies both parameters, IR8 and sSFR, consistently place a galaxy on the main sequence or in the starburst regime. The exception is GN70.8 which is classified as a starburst based on IR8 but lies right on the main sequence given its sSFR. This is a curious exception since, as we discussed, GN70.8 is the close pair system that shows the most convincing evidence for interaction in the optical images as well as showing enhanced SFE from the CO detection. We may be observing this galaxy during a short phase as it transitions from an isolated main sequence galaxy to a starburst as a result of the impending merger. 
Alternatively, the gas fractions in progenitor galaxies can affect these quantities causing the SFE to grow more strongly than SFR in a merger-induced burst (Sargent et al.~2013). 
GN70.104, another close pair, whose CO emission was more localized to one of the optical galaxies, is classified as a main sequence galaxy. 
Overall, we find that half of the 70$\,\mu$m-selected galaxies at $z\sim1$ are designated as starbursts and the other half are main sequence galaxies. 
These independent indicators of the dominant mode of star formation indicate that close pair systems can be designated as either starbursts or main sequence galaxies and that the SFR (and SFE) may be enhanced even at these early stages in a merger. This is consistent with a recent optical morphology study of high redshift ULIRGs which found that a quarter of main sequence galaxies are undergoing interactions indicating external triggering is responsible for some of the stellar mass growth even on the main sequence (Kartaltepe et al.~2012).

\subsection{Linking gas, dust and star formation}
\label{sec:dustgas}

\begin{figure*}
\begin{center}
\includegraphics[width=5.0in,angle=0]{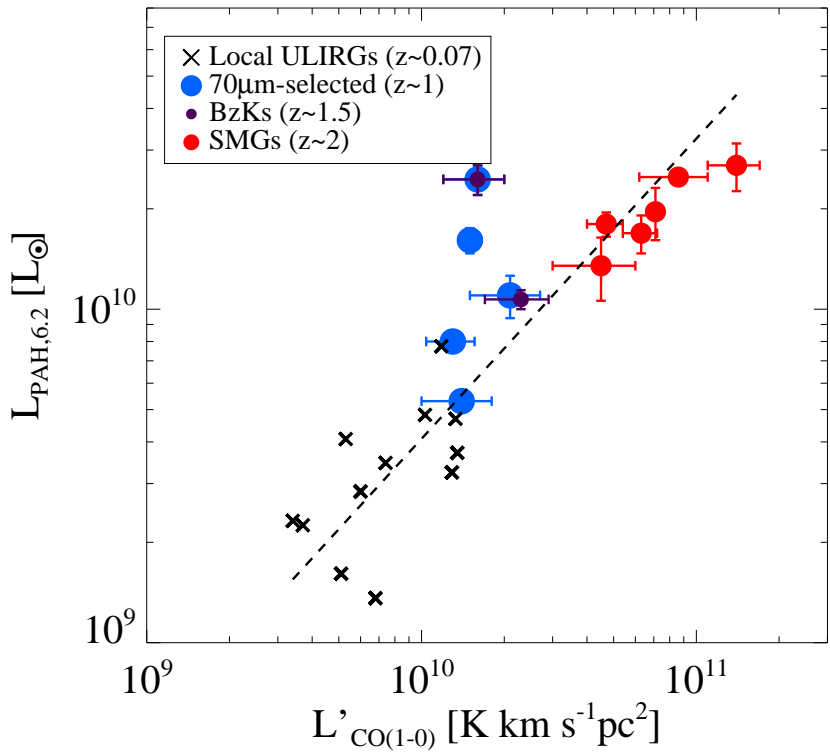}
\caption{Correlation between 6.2$\,\mu$m PAH luminosity and CO luminosity for local and high redshift galaxies. The dashed line shows the best fit to all data points and has a slope of $0.9\pm0.1$. The slightly sub-linear slope may be due to uncertainties in the excitation between upper level CO transitions and CO(1-0) and/or the conversion between CO luminosity and total molecular gas. 
}
\label{fig:lco_lpah}
\end{center}
\end{figure*}

IR, CO and PAH luminosities 
are all used as tracers of the star formation in dusty galaxies although it is clear that the reliability of each tracer is strongly affected by the timescales of star formation. $L^{\prime}_{\rm{CO}}$ measures the current molecular gas reservoir which will fuel future star formation. $L_{\rm{PAH}}$ traces the re-radiated starlight from small dust grains while $L_{\rm{IR}}$ is the sum of the re-radiated emission from all dust grains which may be heated by star formation or AGN activity. 
If star formation is continuously fueled by gas accretion for timescales of about a Gyr then these tracers should all correlate. 
However, when galaxies undergo brief starburst episodes such as those induced by major mergers, the star formation rate can vary significantly over timescales of a few hundred million years (e.g.,~Narayanan et al.~2010). In this case, the timescale for star formation inferred from the molecular gas will be much shorter and the current SFR measured by the dust can not be sustained for long. With an interaction or merger, the AGN can also be fueled which can lead to additional contribution to the IR luminosity.

In local ULIRGs, most of which are ongoing major mergers, the PAH luminosity is observed to decrease as a function of $L_{\rm{IR}}$ possibly as the AGN begins to dominate the mid-IR emission (Desai et al.~2007). Specifically, Tran et al.~(2001) found that most local ULIRGs above $L_{\rm{IR}}>3\times10^{12}L_{\odot}$ are AGN dominated in the mid-IR. It was therefore surprising when high redshift SMGs (with $L_{\rm{IR}}\sim6\times10^{12}L_{\odot}$, Pope et al.~2006) were found to be dominated by PAH emission in the mid-IR with only a small ($\lesssim30\%$) contribution from the AGN (Pope et al.~2008a; Menendez-Delmestre et al.~2009; Coppin et al.~2010). Further IRS observations of high redshift ULIRGs have confirmed that this trend for stronger PAH emission, and correspondingly less mid-IR AGN emission, is not just limited to SMGs and is also seen for many {\it Spitzer}-selected ULIRGs (e.g.,~Pope et al.~2008b; Desai et al.~2009); we are seeing a change in the $L_{\rm{PAH}}$/$L_{\rm{IR}}$ ratio as a function of $L_{\rm{IR}}$ with redshift, at least in the ULIRG regime. 

One fundamental difference between the local Universe and the high redshift Universe is the availability of molecular gas. Cold gas accretion along filaments is much more efficient at high redshift (Kere{\v s} et al.~2009), and indeed observations of molecular gas in isolated galaxies at these redshifts show much higher gas fractions (35--45\%) than local galaxies when assuming a Milky-Way like CO-H$_{\rm{2}}$ conversion factor (Daddi et al.~2010a; Tacconi et al.~2010). 
Recent observational constraints on the conversion factor are consistent with this assumption (Daddi et al.~2010a; Magdis et al.~2012). 
The increased gas fractions could be linked to the enhanced PAH emission seen at high redshift since a significant fraction of the CO and PAH emission originates in PDRs.
In Fig.~\ref{fig:lco_lpah} we plot the 6.2$\,\mu$m PAH luminosity as a function of the CO luminosity for our compilation of high redshift galaxies (Section \ref{sec:hizsample}) and local ULIRGs (Section \ref{sec:localsample}). Despite the scatter, both the local and high redshift galaxies follow a similar correlation between the gas as traced by CO and the dust as traced by PAH emission. The slope of this relationship is slightly sub-linear, $0.9\pm0.1$, indicating that the other factors, such as the CO excitation, the CO-H$_{\rm{2}}$ conversion factor and/or a significant component of CO and/or PAH emission from outside PDRs, may be important.

In Fig.~\ref{fig:lirvratio} we explore the relationship between gas and dust in a different way. 
In the left panel, we plot the ratio of $L_{\rm{PAH,6.2}}$/$L_{\rm{IR}}$ as a function of $L_{\rm{IR}}$. The larger cross symbols show the decreased PAH emission observed in local ULIRGs compared to less luminous local galaxies from 5MUSES (small plus symbols, Wu et al.~2010). The filled colored circles show the high redshift sample including our new observations of 70$\,\mu$m-selected ULIRGs. 
While there is also a trend for $L_{\rm{PAH,6.2}}$/$L_{\rm{IR}}$ to decrease with $L_{\rm{IR}}$ at high redshift, this trend is shifted to higher $L_{\rm{IR}}$ compared to the local Universe. The average $L_{\rm{PAH,6.2}}/L_{\rm{IR}}$ ratio at a given $L_{\rm{IR}}$ is $\sim3$ times higher at high redshift. 

In the middle panel of Fig.~\ref{fig:lirvratio}, we plot $L_{\rm{PAH,6.2}}$/$L_{\rm{IR}}$ again, but as a function of $L_{\rm{IR}}$/$L^{\prime}_{\rm{CO}}$ since we saw in Fig.~\ref{fig:lco_lpah} that the PAH and CO luminosities were correlated. We are limited in the number of galaxies which can be plotted here since we require a clear PAH detection as well as a CO detection, but nevertheless it appears that the high redshift and local galaxies move towards a unified trend of decreasing $L_{\rm{PAH,6.2}}$/$L_{\rm{IR}}$ with increasing $L_{\rm{IR}}$/$L^{\prime}_{\rm{CO}}$, confirming the connection between the PAH and CO emission. Recall that $L_{\rm{IR}}$/$L^{\prime}_{\rm{CO}}$ is a measure of the star formation efficiency (or the inverse of the gas depletion timescale). Within this limited dynamic range of SFE, we find the presence of strong PAH emission to be associated with lower SFE and longer gas depletion times while the weaker PAH galaxies have higher SFE and shorter gas depletion timescales. 

We note that the scatter in the middle panel of Fig.~\ref{fig:lirvratio} is significant. Some of this scatter may be because we have assumed a single excitation to convert from higher order CO transitions down to $L^{\prime}_{\rm{CO(1-0)}}$. In addition, we might expect that the total molecular gas mass is a better tracer of PAH emission than simply the CO emission. 
In the right panel of Fig.~\ref{fig:lirvratio} we show the same plot except with the CO luminosity converted to total molecular gas mass assuming the CO-H$_{\rm{2}}$ conversion factors. While the overall trend persists and might even become tighter, we caution that this last panel is quite uncertain due to a range of acceptable conversion factors.

\section{Discussion}

\begin{figure*}
\begin{center}
\includegraphics[width=6.0in,angle=0]{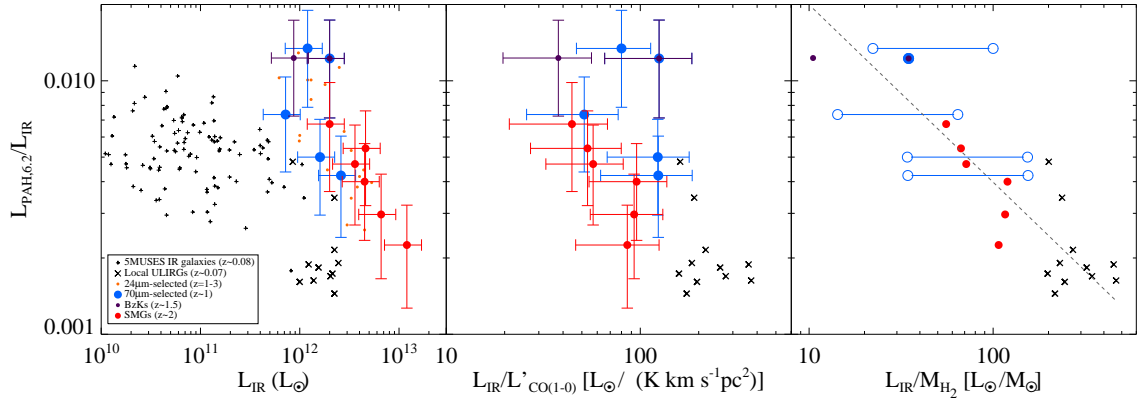}
\caption{PAH line deficiency as a function of $L_{\rm{IR}}$ (left), $L_{\rm{IR}}$/$L^{\prime}_{\rm{CO}}$ (middle) and $L_{\rm{IR}}$/$M_{\rm{H_{2}}}$ (right). The left plot shows a discrepancy between the local and high redshift galaxies: e.g.,~a ULIRG of $2\times10^{12}L_{\odot}$ at $z>1$ is roughly $\sim3$ times brighter in PAH emission than the same ULIRG locally. 
The middle panel shows that after normalizing by the CO luminosity the local and high redshift sources begin to follow the same declining trend.
A very similar trend, namely a deficit in line emission relative to $L_{\rm{IR}}$ shifting to higher $L_{\rm{IR}}$ at high redshift due to increased molecular gas, is seen for the far-IR fine structure lines such as [CII] (GC+11).
The right hand panel shows a variation of the middle plot by converting CO luminosity to total molecular gas mass assuming CO-H2 conversion factors. 
We do not have direct constraints on this value for the 70$\,\mu$m-selected galaxies so they are plotted given a range of acceptable $M_{\rm{H_{2}}}$. Error bars on individual points are omitted from this panel for clarity since we wish to highlight the larger uncertainty from the CO-H$_{\rm{2}}$ for the 70$\,\mu$m-selected galaxies. 
The dashed line is a best fit to all but the 70$\,\mu$m-selected galaxies (slope$\,$=$\,$-0.7, y-intercept$\,$=$\,$-1).
The uncertainty in the CO-H$_{\rm{2}}$ conversion factor limits the interpretation of the last panel. However, our results are suggestive that a relationship exists between the PAH emission and the molecular gas which holds locally and at high redshift. 
}
\label{fig:lirvratio}
\end{center}
\end{figure*}

A very similar trend to what we show in Fig.~\ref{fig:lirvratio} is also seen for [CII] emission; the [CII] deficit observed in local ULIRGs is shifted to higher $L_{\rm{IR}}$ at high redshift (e.g.,~Maiolino et al.~2009; Stacey et al.~2010). 
It is not surprising to find that PAH and [CII] emission show similar trends as they are both often found in PDRs where they trace current star formation (Hollenbach and Tielens 1999; Laurent et al.~2000), although in some lower luminosity galaxies there is a component of PAH and [CII] emission from outside PDRs. 
Within PDRs, PAH emission arises primarily in the warm outer layer closest to the incident UV radiation. [CII] is also found in the inner region and can extend deeper into the PDR to the dissociation front. 
There is a tight correlation observed between PAH and [CII] emission in local star-forming galaxies providing evidence that the atomic interstellar gas is primarily heated by the PAH emission via the photo-electric effect (Helou et al.~2001). CO line emission is found in the cool inner layer of the PDR bounding the molecular cloud (Hollenbach and Tielens 1999; Draine 2011).  
Despite the slightly different regions of the PDR probed by [CII] and CO, there exists strong correlations between CO and [CII] emission from PDRs (e.g.,~Wolfire et al.~1989; Stacey et al.~1991). 
Bendo et al.~(2010) recently made a detailed comparison study of the resolved PAH dust and CO molecular gas in the local star forming galaxy, NGC 2403. While the PAH and CO emission appeared uncorrelated on very small scales ($<330\,$pc), they found very similar radial profiles of the two emission mechanisms in galaxies indicating that globally the processes are linked through the star formation. 

In an attempt to unify the [CII] deficit in high and low redshift galaxies, GC11 found that normalizing $L_{\rm{IR}}$ by the molecular gas brought galaxies at all redshifts onto the same trend. 
We find a similar result where the decrease in $L_{\rm{PAH,6.2}}$/$L_{\rm{IR}}$ occurred at different values of $L_{\rm{IR}}$ at high redshift compared to locally but if we normalize $L_{\rm{IR}}$ by the CO luminosity, this dichotomy is no longer apparent. 
GC11 hypothesize that the galaxies showing decreased $L_{\rm{C[II]}}$/$L_{\rm{IR}}$ (with a corresponding $L_{\rm{FIR}}/M_{\rm{H_{2}}}>80\,$L$_{\odot}$/M$_{\odot}$) are most likely mergers. 
For local star-forming galaxies, an increased radiation field, as expected during a starburst or merger, will cause a decrease in the importance of PAHs in gas heating and dust cooling (Helou et al.~2001). 
In Fig.~\ref{fig:lirvratio}, the sources with the lowest $L_{\rm{PAH,6.2}}$/$L_{\rm{IR}}$, i.e., local ULIRGs and some high redshift SMGs, both show evidence of being major mergers (Sanders et al.~1988; Engel et al.~2010). However we also note that the 70$\,\mu$m-selected galaxies which are close pairs and show SFE=$L_{\rm{IR}}$/$L^{\prime}_{\rm{CO}}$ consistent with mergers have relatively high values of $L_{\rm{PAH,6.2}}$/$L_{\rm{IR}}$. The middle and right panels of Fig.~\ref{fig:lirvratio} show a continuous trend (with significant scatter) and not a clear bimodal distribution of two populations; it may be that at high redshift the turbulent ISM in clumpy disk galaxies can be similar to the ISM conditions induced by mergers or that all galaxies experience mergers of different intensities. It is difficult to distinguish between clumpy gas accretion and minor mergers in current observation and it is possible that most galaxies are affected by both processes.

Another factor which could affect this trend is AGN contribution to one or more of the parameters. 
In the presence of a strong AGN radiation field, PAH molecules can be destroyed depending on their distance from the central AGN (Siebenmorgen et al.~2004). In addition, as the warm continuum component from an AGN begins to dominate in the mid-IR, the observable PAH emission will decrease relative to the total mid-IR luminosity. Of the high redshift sources plotted in Fig.~\ref{fig:lirvratio}, all but GN20 (the lowest SMG point on this figure, Riechers et al.~2013) have a mid-IR AGN fraction $<50\%$.
Furthermore, even a 50\% AGN fraction in the mid-IR typically amounts to a $<10$\% AGN fraction in the total IR luminosity for far-IR/submm selected galaxies (e.g.,~Pope et al.~2008a) which is not enough to drive the trend or much of the scatter seen in Fig.~\ref{fig:lirvratio}. A couple of the local ULIRGs do show a significant AGN in the mid-IR which may boost $L_{\rm{IR}}$ by up to 50\% (Armus et al.~2007). However, if we only plot the portion of the $L_{\rm{IR}}$ from star formation then the local ULIRGs will simply move diagonally toward the top-left corner in the left panel of Fig.~\ref{fig:lirvratio} and they will remain offset from the high redshift galaxies.

A luminous AGN could also act to suppress the CO emission through outflows or harsh radiation fields. Studies of high redshift quasars do not show evidence for decreased CO emission for a given $L_{\rm{IR}}$ compared to other high redshift starburst galaxies (Riechers et al.~2006); however, this result may be due to strong selection effects since the CO-detected quasars are preferentially far-IR luminous.  
The presence of an AGN might also have an effect on the [CII] luminosity, although the continuum emission at $158\,\mu$m is not likely to be coming from the AGN. Nevertheless, Stacey et al.~(2010) do find that the AGN dominated sources in their sample have the largest [CII] deficit relative to $L_{\rm{IR}}$. 
While it is clear that an IR luminous AGN can impact the relations in Fig.~\ref{fig:lirvratio}, larger samples, including more AGN dominated objects with measurable PAH, CO and [CII] emission, are needed to quantify this effect. 

Metallicity can also affect the relative amounts of CO, PAH and [CII] line emission observed in a galaxy. In the local Universe, PAH emission is observed to be suppressed relative to $L_{\rm{IR}}$ in low metallicity galaxies (Draine et al.~2007; Smith et al.~2007). On the other hand, De Breuck et al.~(2011) argue that the enhanced [CII] line emission relative to $L_{\rm{IR}}$ and $L^{\prime}_{\rm{CO}}$ in the highest redshift star-forming galaxies can be attributed to low metallicity. Metallicity can also affect the CO-H$_{2}$ conversion factor if in low metallicity environments most of the carbon is tied up in [CII] instead of CO; this will lead to higher values of $\alpha_{\rm{CO}}$ at low metallicity (Leroy et al.~2011; Genzel et al.~2012).
Taken together these results lead to an anti-correlation between [CII] and PAH emission at low metallicity, although the PAH and CO line emission will still be correlated if both are equally suppressed. The effect of metallicity on these ISM tracers can be tested by combining measurements of PAH, CO, [CII] and metallicity for local and high redshift galaxies when those data exist.

Finally, if the PAH emission accurately traces the small dust grains while the total IR emission traces all dust grains (or the total dust mass, e.g.~da Cunha et al.~2010) then an alternate interpretation of the left panel of Fig.~\ref{fig:lirvratio} is that the small grains play a larger role in photoelectric heating in high redshift galaxies (see also Helou et al.~2001). 
In order to explain the middle and right panels, the small grains would also have to correlate with the molecular gas indicating again that PAH and CO emission are closely linked. 
The hypothesis of changing role of small grains in heating PDRs can be further tested by obtaining [CII] measurements for galaxies with a range of IR and PAH luminosities (e.g.~Helou et al.~2001). 

Regardless of the dominant factor driving the relations in Fig.~\ref{fig:lirvratio}, it remains that the PAH and CO emission in IR luminous galaxies have similar relationships locally and at high redshift. 
If more gas at high redshift correlates with more PAH emission then this could imply that the PDRs are larger for a galaxy of a given luminosity at high redshift. Larger PDRs are present when there is a more diffuse average radiation field and a larger area of star formation. Indeed resolved observations of the gas in high redshift galaxies show much more extended gas reservoirs than local galaxies of the same luminosity (e.g.~Tacconi et al.~2008; Engel et al.~2010; Ivison et al.~2011). 
If PAH emission is the the main source for heating the gas which cools via [CII] emission in PDRs (Helou et al.~2001), then it is expected that they will be correlated and show similar deficits under certain conditions. 
In the future, combining measurements of PAH, CO, [CII] and other far-IR fine structure lines will allow us to piece together a detailed picture of PDRs within the ISM at high redshift and explore how this picture is different for mergers and more continuous star-forming galaxies as function of redshift.

\section{Conclusions}

The paper provides the first comparison of PAH dust emission and molecular gas, as traced by CO, in high redshift galaxies. Combining our new measurements with those from the literature, we study a sample of 12 high redshift galaxies which have both PAH and CO detections and compare these to a sample of 14 local ULIRGs. 
We present new CO and/or PAH measurements for 6 ULIRGs at $z\sim1$, selected based on their 70$\,\mu$m emission.
We find that the counterparts to the CO detected 70$\,\mu$m ULIRGs are close pairs of two optical galaxies with consistent spectroscopic redshifts indicating an early stage of a merger or interaction. 
We find high ratios of $L_{\rm{IR}}$/$L^{\prime}_{\rm{CO}}$, a commonly used proxy for the SFE, consistent with values found for major mergers locally and at high redshift.
Whether or not the IR and/or CO emission are affected by the early stages of an interaction requires higher resolution submm and CO observations to determine how the gas and dust is distributed in these complex systems. 
Comparing the sSFR and the ratio of $L_{\rm{IR}}/L_{\rm{8\mu m}}$ to those found for typical star-forming galaxies at $z\sim1$, we find that the close pair systems can either live on the main sequence or have elevated SFRs; the SFR may be externally triggered even at these early stages in a merger enhancing the SFE. 

These 70$\,\mu$m-selected sources are all rich in PAH emission indicating little or no AGN contribution to the mid-IR luminosity. 
This is an interesting comparison sample to the high redshift SMGs, which also show a small AGN contribution in the mid-IR. SMGs are selected based on their cold dust emission while the 70$\,\mu$m selection can act to preferentially pick up galaxies with warmer dust emission. 
Large comprehensive samples are need to fully understand the role of AGN activity in the mid-IR as a function of IR luminosity and dust temperature in high redshift galaxies.
We measure the PAH luminosities in individual galaxies as a tracer of the SFR through PDRs. 

Given what we have learnt from detailed studies of local star-forming galaxies, we explore the PAH luminosity as a parameter for constraining the conditions of the ISM in high redshift IR luminous galaxies. 
We show that the average $L_{\rm{PAH,6.2}}/L_{\rm{IR}}$ ratio at a given $L_{\rm{IR}}$ is $\sim3$ times higher at high redshift. 
The decrease in $L_{\rm{PAH,6.2}}/L_{\rm{IR}}$ ratio as a function of $L_{\rm{IR}}$ occurs at higher luminosities for high redshift galaxies, similar to what has been found for [CII] and other far-IR lines tracing PDRs. 
Once the IR luminosity is normalized by the molecular gas we find the local and high redshift ULIRG follow a similar decline in $L_{\rm{PAH,6.2}}$/$L_{\rm{IR}}$ as a function of $L_{\rm{IR}}$/$L^{\prime}_{\rm{CO}}$. 
In future studies, measurements of CO, [CII] and PAH emission can be combined to develop a detailed picture of the ISM in high redshift galaxies and how it is affected by major mergers and continuous gas accretion. 
A detailed comparison study of spatially resolved PAH, CO and [CII] emission in local galaxies will provide a benchmark for interpreting studies of the ISM at high redshift. We look forward to future observations with the Atacama Large Millimeter Array (ALMA) and the James Webb Space Telescope (JWST) which will allow us to probe all of these ISM tracers in larger samples of high redshift galaxies.

\acknowledgments
We thank the referee for their detailed comments which improved the clarity of this work. We thank Sascha Trippe for his help with the IRAM PdBI data reduction and the IRAM staff for carrying out the observations. 
We are grateful to Yong Shi for providing the data points for local galaxies from Wu et al.~(2010). 
AP thanks Desika Narayanan and Jeyhan Kartaltepe for helpful discussions.
This work is based in part on observations made with the {\it Spitzer Space Telescope}, which is operated by the Jet Propulsion Laboratory, California Institute of Technology under a contract with NASA, and the {\it Herschel Space Observatory}, a European Space Agency Cornerstone Mission with significant participation by NASA. Support for this work was provided by NASA through an award issued by JPL/Caltech.

\clearpage

\begin{table*}
\begin{center}
\scriptsize
\caption{Targets with {\it Spitzer} IRS and/or IRAM PdBI observations}
\label{tab:sample}
\begin{tabular}{lllllllll}
\hline
ID & RA  & DEC & z$_{\rm{optical}}$$^{\rm{a}}$ & $S_{24}$ & $S_{70}$ & $L_{\rm{IR}}$$^{\rm{b}}$  & IR8$\,=L_{\rm{IR}}/L_{8\mu m}$$^{\rm{c}}$  & sSFR=SFR/M$_{\ast}$$^{\rm{c}}$\\
& (J2000)  & (J2000) &   &  (mJy)  & (mJy) & ($10^{12}\,L_{\odot}$) & & ($10^{-9}\,$yr$^{-1}$) \\\hline
GN70.38       &   12:36:33.68  & 	+62:10:05.9  & 	1.0156 & $0.572\pm0.008$ &  $2.7\pm0.7$ & $0.72\pm0.29$ & 5.8 (MS) & 0.49 (MS)\\ 
GN70.8         &   12:36:20.94  & 	+62:07:14.2  & 	1.1482/1.1497 & $0.447\pm0.010$ &  $5.3\pm0.7$ &  $1.6\pm0.6$ & 9.1 (SB) & 0.72 (MS)\\
GN70.104    &    12:37:02.74   & 	+62:14:01.5 & 	1.2463/1.2435 & $0.344\pm0.008$ &  $2.4\pm0.7$	 &  $1.2\pm0.5$ & 5.4 (MS) & 1.1 (MS)\\
GN70.61      &    12:36:53.37  & 	+62:11:39.6  & 	1.2698 & $0.339\pm0.008$ &  $7.0\pm0.7$ & $2.2\pm0.9$ & 11.4 (SB) & 4.4 (SB)\\
GN70.14       &   12:36:45.83  & 	+62:07:54.2  & 	1.4320/1.4340 & $0.366\pm0.006$ &  $5.6\pm0.7$ & $2.6\pm1.0$ & 10.4 (SB) &  2.1 (SB)\\
GN70.211     &   12:37:10.62   & 	+62:22:34.5  & 	1.5230 & $0.383\pm0.007$ &  $4.9\pm0.7$ & $2.0\pm0.8$ & 6.5 (MS) & 2.1 (MS)\\
\hline
\end{tabular}
\end{center}
$^{\rm{a}}$\,Optical spectroscopic redshifts from Wirth et al.~(2004) and Barger et al.~(2008).  \\
$^{\rm{b}}$\,Errors on $L_{\rm{IR}}$ are conservatively estimated to be 40\% which includes uncertainty due to the scatter in the SED shapes and the photometric uncertainties (see Kirkpatrick et al.~2012). \\
$^{\rm{c}}$\,Main Sequence (MS) and StarBurst (SB) designations based on main sequence relations given in Equations (2) and (14) of Elbaz et al.~(2011). 
\end{table*}

\begin{table*}
\begin{center}
\scriptsize
\caption{IRAM PdBI measurements 
}
\label{tab:co}
\begin{tabular}{lllllllll}
\hline 
ID & T$_{\rm{int}}$ & z$_{\rm{CO}}$ & FWHM & $I_{\rm{CO(2-1)}}$ & $L^{\prime}_{\rm{CO(2-1)}}$  &$L^{\prime}_{\rm{CO(1-0)}}$$^{\rm{a}}$  & $M_{\rm{H_{2}}}$$^{\rm{b}}$ \\
&  (hr) &  &  (km$\,$s$^{-1}$) & (Jy km/s) & $(10^{10}\rm{K\,km\,s^{-1}pc^{2}}$)   & $(10^{10}\rm{K\,km\,s^{-1}pc^{2}}$)    & $(10^{10}\,$M$_{\odot}$)   \\\hline
GN70.8 & 7.4 & 1.1489 & $430\pm129$  &  $0.57\pm0.13$ & $1.0\pm0.2$  &  $1.3\pm0.3$ &  1.0--4.7  \\
GN70.104  & 9.9 & 1.2463 & $139\pm40$ & $0.57\pm0.05$ &  $1.2\pm0.1$    & $1.5\pm0.2$ &  1.2--5.4\\
 \hline
\end{tabular}
\end{center}
$^{\rm{a}}$\,Assuming $r_{21}=0.8\pm0.1$ (Aravena et al.~2010; Frayer et al.~2011).
$^{\rm{b}}$\,Assuming a range of CO-H$_{\rm{2}}$ conversion factors: 0.8--3.6$\,\rm{M_{\odot}\,(K\,km\,s^{-1}pc^{2})^{-1}}$ (Tacconi et al.~2008; Daddi et al.~2010a; Magdis et al.~2012).
\end{table*}

\begin{table*}
\begin{center}
\scriptsize
\caption{{\it Spitzer} IRS observations and measurements
}
\label{tab:pah}
\begin{tabular}{lllllll}
\hline
ID   & \multicolumn{3}{c}{Integration time}  & Mid-IR continuum$^{\rm{a}}$ &  $L_{\rm{PAH,6.2}}$    \\
&  SL1 ($\times$240s) & LL2 ($\times$120s) & LL1 ($\times$120s)  & (\%)  & ($10^{9}\rm{L_{\odot}}$)  \\\hline
GN70.38 & 9$\times$2   & 6$\times$2 & 10$\times$2 & $<35$ & $5.3\pm0.2$      \\  
GN70.8 &   8$\times$2   & 6$\times$2  & 10$\times$2 & $<35$  &  $8.0\pm0.5$      \\
GN70.104 & 5$\times$2 & 4$\times$2 & 11$\times$2 & $<35$ &  $16.1\pm1.4$      \\
GN70.61   & 5$\times$2 & 4$\times$2 &  12$\times$2 & $24\pm12$ &  $10.0\pm0.8$      \\
GN70.14   & &  6$\times$2 & 10$\times$2  & $<35$ &  $11.0\pm1.6$       \\
GN70.211 &    & 7$\times$2 & 10$\times$2  & $<35$ &  $24.5\pm2.5$      \\
 \hline
\end{tabular}
\end{center}
$^{\rm{a}}$\,Percentage of mid-IR flux that comes from continuum emission in spectral decomposition.  
\end{table*}

\begin{table*}
\begin{center}
\scriptsize
\caption{Compilation of CO and PAH measurements for high redshift ULIRGs (including values from the literature). 
}
\label{tab:hiz}
\begin{tabular}{lllllllll}
\hline
ID & Type & RA & DEC & Redshift & $L^{\prime}_{\rm{CO}}$ & $L_{\rm{PAH,6.2}}$ & $L_{\rm{IR}}$$^{\rm{b}}$  & References$^{\rm{c}}$ \\
& &   & & & $(10^{10}\rm{K\,km\,s^{-1}pc^{2}}$) & ($10^{10}\rm{L_{\odot}}$)  & ($10^{12}\rm{L_{\odot}}$)  & [CO,PAH] \\\hline
GN70.38    & 70$\mu$m    &   12:36:33.68  & 	+62:10:05.9  & 	1.016 & $1.1\pm0.3$ (2-1) & $0.53\pm0.02$   &  $0.72\pm0.29$  &  M12,this paper  \\ 
GN70.8 & 70$\mu$m &  12:36:20.94  & 	+62:07:14.2  &  1.148 &    $1.0\pm0.2$ (2-1)  & $0.80\pm0.05$ & $1.6\pm0.6$ & this paper  \\
GN26  &  SMG/70$\mu$m & 12:36:34.51 &  +62:12:40.9 &    1.223 &    $6.9\pm1.9$ (2-1)  & $2.49\pm0.06$ &   $4.6\pm1.8$ & F08,P08 \\
GN70.104 & 70$\mu$m & 12:37:02.74   & 	+62:14:01.5  &  1.246  &    $1.2\pm0.1$ (2-1)  & $1.61\pm0.14$ & $1.2\pm0.5$  & this paper  \\
GN70.14 & 70$\mu$m &  12:36:45.83  & 	+62:07:54.2  &  1.432  &    $1.7\pm0.5$ (2-1)  & $1.10\pm0.16$ & $2.6\pm1.0$ & C11,this paper  \\
BzK4171 & BzK & 12:36:26.53 & +62:08:35.3 & 1.465   &    $2.3\pm0.6$ (1-0)  & $1.07\pm0.07$ & $0.87\pm0.35$ & A10,this paper  \\
GN70.211 (BzK21000) & 70$\mu$m,BzK & 12:37:10.60 & +62:22:34.6 &  1.523   &    $1.6\pm0.4$ (1-0)  & $2.45\pm0.25$  & $2.0\pm0.8$  & A10,this paper  \\
GN39b & SMG & 12:37:11.97 & +62:13:25.8 & 1.992   &     $2.7\pm0.9$ (3-2)   & $1.35\pm0.29$ &  $2.0\pm0.8$ &  B13,P08 \\
GN39a & SMG & 12:37:11.37 & +62:13:31.1 & 1.996   &   $4.2\pm0.1$ (3-2)   & $1.96\pm0.35$ &  $6.6\pm2.6$ & B13,P08 \\
GN06 & SMG & 12:36:18.33 & +62:15:50.4 & 2.000   &   $1.9\pm0.3$ (4-3)   & $1.80\pm0.15$ &  $4.5\pm1.8$ & B13,P08 \\
GN19$^{\rm{a}}$ & SMG &12:37:07.19 &   +62:14:08.0 &   2.490 &    $6.3\pm0.9$ (1-0)    & $1.69\pm0.22$ & $3.6\pm1.4$  & I11,P08 \\
GN20  & SMG & 12:37:11.86 & +62:22:11.9 &  4.055  &   $14\pm3$ (1-0)  & $2.70\pm0.44$ &  $12\pm5$ & C10,R13 \\
\hline
\end{tabular}
\end{center}
$^{\rm{a}}$\,This SMG has two CO counterparts separated by 3 arcsec. Both are within the IRS beam and so we combine the CO measurements to compare the gas and dust for the whole system. \\
$^{\rm{b}}$\,Errors on $L_{\rm{IR}}$ are conservatively estimated to be 40\% which includes uncertainty due to the scatter in the SED shapes and the photometric uncertainties (see Kirkpatrick et al.~2012). \\
$^{\rm{c}}$\,M12: Magnelli et al.~(2012), F08: Frayer et al.~(2008), P08: Pope et al.~(2008), C11: Casey et al.~(2011), A10: Aravena et al~(2010), B13: Bothwell et al.~(2013), I11: Ivison et al.~(2011), C10: Carilli et al.~(2010), R13: Riechers et al.~(2013).
\\
\end{table*}

\begin{table*}
\begin{center}
\scriptsize
\caption{Local ULIRG sample; the first 7 are from Armus et al.~(2007) and the rest are selected to be strong PAH sources from Desai et al.~(2007).
}
\label{tab:local}
\begin{tabular}{lllll}
\hline
ID & Redshift & $L^{\prime}_{\rm{CO(1-0)}}$ & $L_{\rm{PAH,6.2}}$ & $L_{\rm{IR}}$$^{\rm{h}}$  \\
&   & $(10^{9}\rm{K\,km\,s^{-1}pc^{2}}$) & ($10^{8}\rm{L_{\odot}}$)  & ($10^{12}\rm{L_{\odot}}$)   \\\hline
Arp220  & 0.018 & $7.1\pm1.4$$^{\rm{a}}$  & $5.2\pm0.1$   &1.4$^{\rm{e}}$  \\
IRAS 05189-2524   & 0.042 &  $7.6$$^{\rm{b}}$ & $4.9\pm0.8$  & 1.4$^{\rm{e}}$   \\
Mrk 273 & 0.038 & $6.8\pm0.3$$^{\rm{c}}$  & $13.6\pm0.4$  & 1.4$^{\rm{e}}$  \\
UGC 5101 & 0.039  &   $5.1\pm0.2$$^{\rm{c}}$   & $16.1\pm0.4$  & 1.0$^{\rm{e}}$  \\
IRAS 22491-1808 & 0.077  & $6.0$$^{\rm{b}}$  &  $28.4\pm0.4$  & 1.5$^{\rm{e}}$  \\
IRAS 12112+0305 & 0.073 & $13.5\pm1.5$$^{\rm{d}}$ &   $37.1\pm0.4$   & 2.1$^{\rm{e}}$  \\
IRAS 14348-1445 & 0.083  & $12.9$$^{\rm{b}}$  &  $32.4\pm0.5$    & 2.2$^{\rm{e}}$  \\ 
\hline
IRAS 20414-1651 & 0.086 & $3.7\pm1.3$$^{\rm{d}}$  & $22.5\pm0.1$   & 1.7$^{\rm{f}}$ \\
IRAS 19458+0944 & 0.100 &  $13.3\pm2.7$$^{\rm{a}}$ &  $46.9\pm0.2$   & 2.5 $^{\rm{f}}$ \\ 
IRAS 22491-1808 & 0.077 & $3.4\pm0.8$$^{\rm{d}}$ &  $23.2\pm0.2$    & 1.4 $^{\rm{f}}$ \\
IRAS 19297-0406 & 0.085 & $7.4\pm1.4$$^{\rm{d}}$ &   $34.6\pm0.3$   & 2.6 $^{\rm{f}}$ \\
IRAS 10565+2448 & 0.043 & $5.3\pm0.4$$^{\rm{d}}$ &   $40.8\pm0.2$   & 1.1 $^{\rm{f}}$ \\
IRAS 17208-0014 & 0.043 & $10.3\pm0.6$$^{\rm{d}}$ &   $48.2\pm0.2$   & 0.98 $^{\rm{f}}$ \\
IRAS 16334+4630 & 0.191 & $11.8\pm4.1$$^{\rm{a}}$ &   $77.4\pm0.8$   & 2.2 $^{\rm{g}}$\\
 \hline
\end{tabular}
\end{center}
$^{\rm{a}}$\,Solomon et al.~1997, 
$^{\rm{b}}$\,Sanders et al.~1991 (no uncertainties given),
$^{\rm{c}}$\,Gao \& Solomon 2004, 
$^{\rm{d}}$\,Chung et al.~2009,
$^{\rm{e}}$\,Armus et al.~2007,
$^{\rm{f}}$\,Farrah et al.~2003,
$^{\rm{g}}$\,Kim, Veilleux \& Sanders 1998
$^{\rm{h}}$\, Uncertainties on $L_{\rm{IR}}$ are typically $\lesssim5\%$ (U et al.~2012). 
\end{table*}

\end{document}